\documentclass[epjH]{svjour}
\usepackage{epsfig}
\usepackage{graphicx}
\usepackage {amsmath}
\usepackage {amssymb}
\usepackage {bm}  
\usepackage{latexsym}
\input {colordvi}
\def\lsim{\raise0.3ex\hbox{$<$\kern-0.75em\raise-1.1ex\hbox{$\sim$}}}
\def\gsim{\raise0.3ex\hbox{$>$\kern-0.75em\raise-1.1ex\hbox{$\sim$}}}

\def\Journal#1#2#3#4{{#4}. {\it #1} {\bf #2}: #3}
\def\IJMPA{{Int. J. Mod. Phys. A}}
\def\EPJC{{Eur. Phys. J. C}}

\def\NCA{Nuovo Cimento\ }

\def\NIMA{{Nucl. Instrum. Methods A}}
\def\NPA{{Nucl. Phys. A}}
\def\NPB{{Nucl. Phys. B}}
\def\PLB{{Phys. Lett. B}}
\def\PLC{Phys. Repts.\ }
\def\PL{Phys. Lett.\ }
\def\PRL{Phys. Rev. Lett.\ }
\def\PRD{{Phys. Rev. D}}

\def\PR{Phys. Rev.\ }

\def\ARNPS{{Ann. Rev. Nucl. Part. Sci.\ }} 

\def\PHP{Phys. Perspect.\ }
\def\HIP{Heavy Ion Phys.\ }

\begin{document}
\title{Waiting for the $W$}
\subtitle{and the Higgs}
\author{M.~J.~Tannenbaum\thanks{\email{mjt@bnl.gov}}}
%
\institute{Physics Department, Brookhaven National Laboratory Upton, NY 11973-5000 USA \ }
\abstract{The search for the left-handed $W^{\pm}$ bosons, the proposed quanta of the weak interaction, and the Higgs boson, which spontaneously breaks the symmetry of unification of electromagnetic and weak interactions, has driven elementary-particle physics research from the time that I entered college to the present and has led to many unexpected and exciting discoveries which revolutionized our view of subnuclear physics over that period. In this article I describe how these searches and discoveries have intertwined with my own career. }
\maketitle



\section{Introduction}
\subsection{Columbia College---1955-1959}
When I was a sophomore at Columbia College, my physics teacher, Polykarp Kusch, had just won the Nobel Prize. In my senior year, another teacher, Tsung-Dao (T.D.) Lee, had also just won the Nobel Prize. I thought this was normal. 

Kusch together with Willis Lamb (then at Yale) were awarded the Prize for two experiments which proved the validity and precision of Quantum Electrodynamics: the anomalous magnetic moment of electron~\cite{KuschPR73} ($g_e - 2$); and the (Lamb) shift of two fine structure levels in Hydrogen~\cite{Lamb47}. Both these experiments were performed at Columbia a few stories higher in Pupin Laboratories than the classrooms where I attended College and Graduate School. 

Lee (and Yang) were awarded the Prize for the suggestion of Parity Violation in the Weak Interaction ($\beta$ decay)~\cite{LY56} which was discovered by C.~S.~Wu, another Columbia professor, and collaborators~\cite{WuPV57} in the $\beta$ decay of polarized Cobalt$^{60}$ and confirmed via the Columbia grapevine a few days later  at Columbia's Nevis Cyclotron in the $\pi^+\rightarrow \mu^+ +\nu$, $\mu^+\rightarrow e^+ +\nu +\bar{\nu}$ decay chain~\cite{GLW57} which also produced the first measurement of the free $\mu^+$ magnetic moment $g_{\mu^+}=+2.00\pm 0.10$. 

   It should come as no surprise that I only applied to one graduate school. 
     
\subsection{Graduate School at Columbia---1959-1965}
	In early 1960, Mel Schwartz~\cite{MS60} and Lee and Yang~\cite{LY60} discussed the feasibility of using high energy neutrinos to study the weak interactions. A few months later Lee and Yang proposed the intermediate bosons $W^{\pm}$ as the quanta that transmit the weak interaction~\cite{LYW60} and Mel Schwartz proposed to detect them with a beam of high energy neutrinos at the new Alternating Gradient Synchrotron (AGS) of Brookhaven National Laboratory (BNL), experiment AGS-28, proposed May 26,1960~\cite{MelBio}. Later that year Mel asked me to be a thesis student on this experiment;  but I turned him down because there were already 2 thesis students on the experiment. 
	
\begin{figure}[!thb]
\begin{center}
\includegraphics[width=0.95\linewidth]{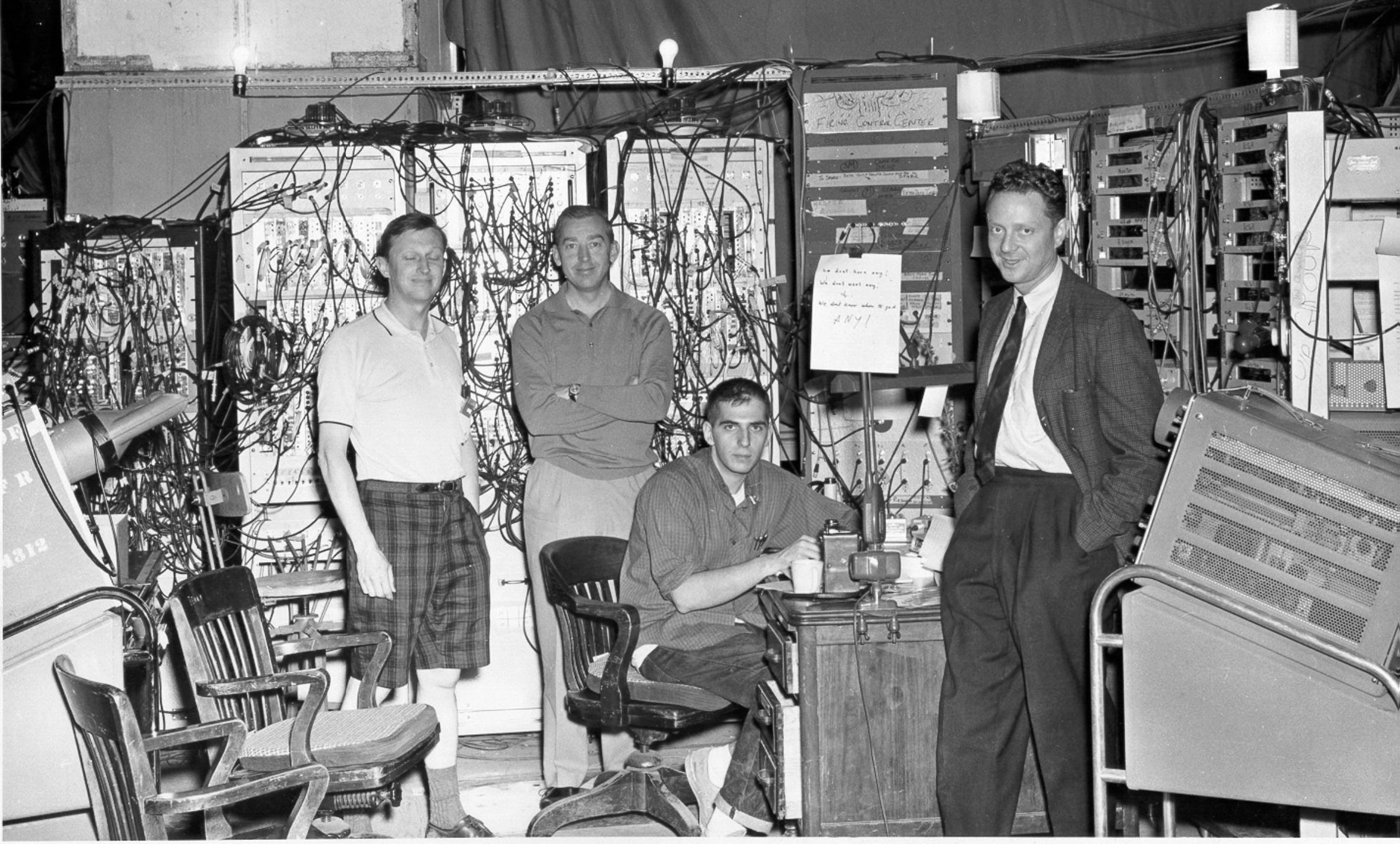}
\end{center}
\caption[]{Bosses of $\mu$-p I experiment, c. 1961. (Standing-Left to Right) John Tinlot, Rod Cool, Leon Lederman. Seated is the graduate student, Michael Tannenbaum. }
\label{fig:bosses}\vspace*{-0.12in}
\end{figure}

	I chose to work with Leon Lederman in his search for the force of muness---``why  does the muon weigh heavy?'' I guess that one of the reasons I decided to go with Leon instead of with Mel is that I had spent most of the summer of 1960 at BNL talking to Leon about this problem when I was a summer student of John Russell and Satoshi Ozaki working at the Cosmotron. I spent lots of time in the Cosmotron library reading everything there was to know about electron elastic and inelastic scattering, which was really a hot topic at that time. My thesis was experiment $\mu$-p I  (AGS-4), muon proton elastic scattering  (Fig.~\ref{fig:bosses}), which compared the radius of the proton measured with muons to that of Robert Hofstadter's and Bob Wilson's measurements with electrons, to deduce the difference in the radii of electrons and muons probed by hitting a proton. There was no effect~\cite{mu-p-I}, the proton form factor measured with both muons and electrons was identical  all the way out to $Q^2=1$ GeV$^2$/c$^2$ \mbox{[25.7 inverse fm$^2$]}. (Never did I realize that learning these units would help me in the 21st century at RHIC---the BNL Relativistic Heavy Ion Collider). 

	The results of this measurement were first presented by Leon and John Tinlot (of Rochester)~\cite{Tinlot} at the 12th International Conference on High Energy Physics which was held in Dubna, Russia in 1964 (a really big deal at the time) so I pored over the proceedings of this conference to see how the result was received. This was the time when questions and discussion as well as the text of the talks were included in the proceedings, and a good question was asked by an A. (Antonino) Zichichi. As I thumbed through the proceedings I found another good comment from Zichichi~\cite{Zichichi-Dubna} about how to find the $W$ boson in p+p collisions, which happens to be the way the $W$ was discovered 19 years later. It is so beautiful and concise that I quote it verbatim: ``We would observe the $\mu$'s from W-decays. By measuring the angular and momentum distribution at large angles of K and $\pi$'s, we can predict the corresponding $\mu$-spectrum. We then see if the $\mu$'s found at large angles agree with or exceed the expected numbers. A supplementary check can be made by measuring the polarization of these $\mu$'s. The polarization indicates the origin of these $\mu$'s.''
\section{CERN 1965-1966}
    After graduate school, I went to CERN as a post-doc in 1965 and worked with Jack Steinberger and Carlo Rubbia on the first $K^0_S$ and $K^0_L\rightarrow \pi^+ \pi^-$ interference experiment~\cite{Alff} following the discovery of CP violation in $K^0_L$ decay at BNL~\cite{FitchCronin} during my thesis run. I was actually on shift at both the BNL-AGS and the CERN-PS when the motor generators (MG) died in both places during this  period. The announcement over the loudspeaker at the AGS was something like:``The beam is off, the beam is off, all targets have stopped flipping, all targets have stopped flipping. The beam will be off for approximately 6 months.'' The 6 month delay in early 1966 at CERN caused by the MG failure  resulted in the rescheduling of Jack and Carlo's  follow-up $K^0$ ``vacuum regeneration'' experiment, so I switched over to work on the second muon $g-2$ experiment, the first one that used a storage ring. The original CERN muon $g-2$ experiment~\cite{Charpak1961} which used a 6m long tapered dipole was CERN's main claim to fame during this era. 
    
    I had many discussions with Francis Farley and Emilio Picasso (and Zichichi) in this period on the structure of the proton and on how muons and electrons could be different. I then went to Harvard as an Assistant Professor (eventually Associate Professor) and continued thinking about the ``force (or quantum) of $\mu$-ness''. One of the reasons that I and others went to Harvard at this time was the Cambridge Electron Accelerator (CEA)---located close to the museum with the glass flowers on Oxford Street---at which the ``Pipkin Effect'', the anomalous photo-production of wide angle $e^+ e^-$ pairs had been discovered in 1965~\cite{PipkinEffect}. Emilio and I got Norman Kroll (another of my Columbia teachers, who was visiting CERN) to explain to us for the paper from the $g-2$ measurement~\cite{g-2} why this apparent violation of QED  could only be due to a violation of gauge invariance or, as suggested by Francis Low~\cite{Low1965}, the production of an excited electron, $e^*\rightarrow e+\gamma$. There was also another possibility~\cite{Ting67}.~\footnotemark 
\footnotetext{The Pipkin Effect~\cite{PipkinEffect} was wrong because they essentially divided by zero by making the measurement for exactly symmetric pairs, where the cross-section is identically zero by gauge invariance. Their detector Monte Carlo didn't miss by much but I don't think that they realized that the cross section they were trying to measure was exactly zero. However this wrong result spurred lots of good theoretical ideas. Sam Ting first became famous by going to DESY and showing that QED was indeed correct for this reaction~\cite{Ting67}. He had some help from Stan Brodsky, also at Columbia at the time.}
\section{Harvard 1966-1971}
    I arrived at Harvard in the fall of 1966 at the same time as Steve Weinberg who was visiting as the Morris Loeb Lecturer. Steve then moved to MIT in the fall of 1967 where he wrote his seminal ``A model of leptons'' paper~\cite{Weinberg-leptons} in which the symmetries of the weak and electromagnetic interactions are unified but spontaneously broken by a scalar boson leading to observable states of a massless photon $\gamma$ which mediates the electromagnetic interaction, the massive $W^{\pm}$ bosons and a new neutral $Z^0$ boson that mediate the weak interaction as well as a new scalar Higgs boson. 
    
    My story during this 5 year period was somewhat different.
    
	Almost as soon as I arrived at Harvard, Leon Lederman asked me to collaborate on a review article on muon physics~\cite{LML-MJT} that he had been commissioned to write. I got two interesting new ideas in writing this article: the formalism for the design of transversely and longitudinally polarized muon beams (which was actually used by the Spin Muon Collaboration~\cite{SMC}); and the idea to measure the force between muons by scattering muons from each other---the analogy being the strong force which is not visible in muon-proton scattering but knocks you over in p+p scattering. Although a hot-topic at BNL now, colliding beams of muons didn't exist at that time and still don't exist, so I had the idea of using a muon to produce a muon-pair in the field of a nucleus, the muon-trident, $\mu^{\pm} + Z\rightarrow \mu^{\pm} + Z +\mu^+ + \mu^-$. Also, since there would be two-identical muons in the final state, this reaction would measure the statistics of the muon---with para-statistics such as the existence of red, green  and blue muons~\cite{Greenberg64} (sound familiar?) being a possible explanation of the $\mu$-e difference.  
	
	In order to calculate the rates and sensitivities for the proposal to the AGS, I needed the muon-trident cross sections. Stan Brodsky and Sam Ting (both of whom I had known at Columbia while a graduate student) had just written a long article on this process as a test of QED~\cite{Brodsky-Ting} but had only shown differential distributions in the 3 independent solid angles and two energies while I needed the integrated cross section. Stan had computed all the effects including the exchange diagrams for identical muons in the final state and kindly sent me his computer code for calculating the matrix elements, which was a full box ($\sim 2000$) of IBM cards. I succeeded~\cite{MJT-tridents} in summing, squaring and integrating the matrix elements using a novel  Monte Carlo technique which was: ``a tricky business because there was no simple analytical expression to use as a guide'' and because the matrix elements diverged any time any of the 6 momenta involved became collinear. This exercise started a long-lasting relationship of common interests and friendship between Stan and me, resulting in, among other things, Stan's proposal to use direct-pair production by anti-protons followed by capture of the positron as the first method to make anti-hydrogen~\cite{Munger} soon after my having invited him to BNL in 1990 to a workshop on ``Can RHIC be used to test QED''~\cite{RHIC-QED} at which he provided the famous 34kb cross section for Au+Au $\rightarrow$ Au+Au + $e^+ +e^-$ at $\sqrt{s_{NN}}=200$ GeV (which followed by the capture of the $e^-$ by an Au$^{79{^+}}$ nucleus and its subsequent loss from the beam due to the change in charge is the limitation of the  luminosity in Au+Au, or Pb+Pb,  colliders).  

	Coming back to the muon-trident experiment, it was approved and ran for 200.0 hours at the AGS in January-February 1968 with the result of Fermi statistics for the muon and no anomalous muon-muon force~\cite{Trident-PRL}. Interestingly, the first preliminary results from this experiment were presented by Pief Panofsky as a rapporteur at the 14th ICHEP in Vienna in 1968~\cite{Pief-Vienna} and I was a bit miffed at the time because I thought that he didn't give it enough space compared to the new SLAC result on the ``continuum excitation'' in inelastic electron scattering, the now famous Deeply Inelastic Scattering. In retrospect, Pief's talk was really great, and fair to the large number of other results whose  significance pales in comparison to the DIS discovery. 

	The Bjorken scaling~\cite{BJscaling}, which led to the understanding of the DIS measurement, together with the results of another di-muon experiment at the AGS in this period (the discovery of Drell-Yan muon pairs~\cite{Christenson1970}) led Leon Lederman to propose in the ISABELLE workshop series of 1971-72 that to detect the violation of unitarity of the pure weak interaction $p+p\rightarrow \mu + \nu_{\mu} +X$ (if there were no $W^{\pm}$), or to detect the $W^{\pm}$  at ISABELLE (p+p collider, $\sqrt{s}=400$ GeV), ``the luminosity required is $\gsim 5 \times 10^{33}$ cm$^{-2}$ s$^{-1}$. At this luminosity, ISABELLE is guaranteed to make at least one fundamental discovery!''~\cite{Lederman1972} You might ask why Leon was looking at di-muons in p+p collisions, which had no relation to muon-muon scattering. Actually Bob Crease covered this nicely in his articles on ``The ISABELLE saga''~\cite{Crease-saga}. In brief, Leon and others who were looking for $W^{\pm}$ production in p+p collisions needed to know the time-like form factor of the proton, which could be measured by di-lepton production ($\bar{p}+p\rightarrow \mu^+ +\mu^-$ or \mbox{$p+p\rightarrow \mu^+ +\mu^- +X$})~\cite{Chilton1966,Yamaguchi1966}, in order to set limits on the $W^{\pm}$ cross section (and thus mass) from their non-observation of any excess high $p_T$ single leptons.  

\begin{figure}[!b]
\begin{center}
\includegraphics[width=0.7\linewidth]{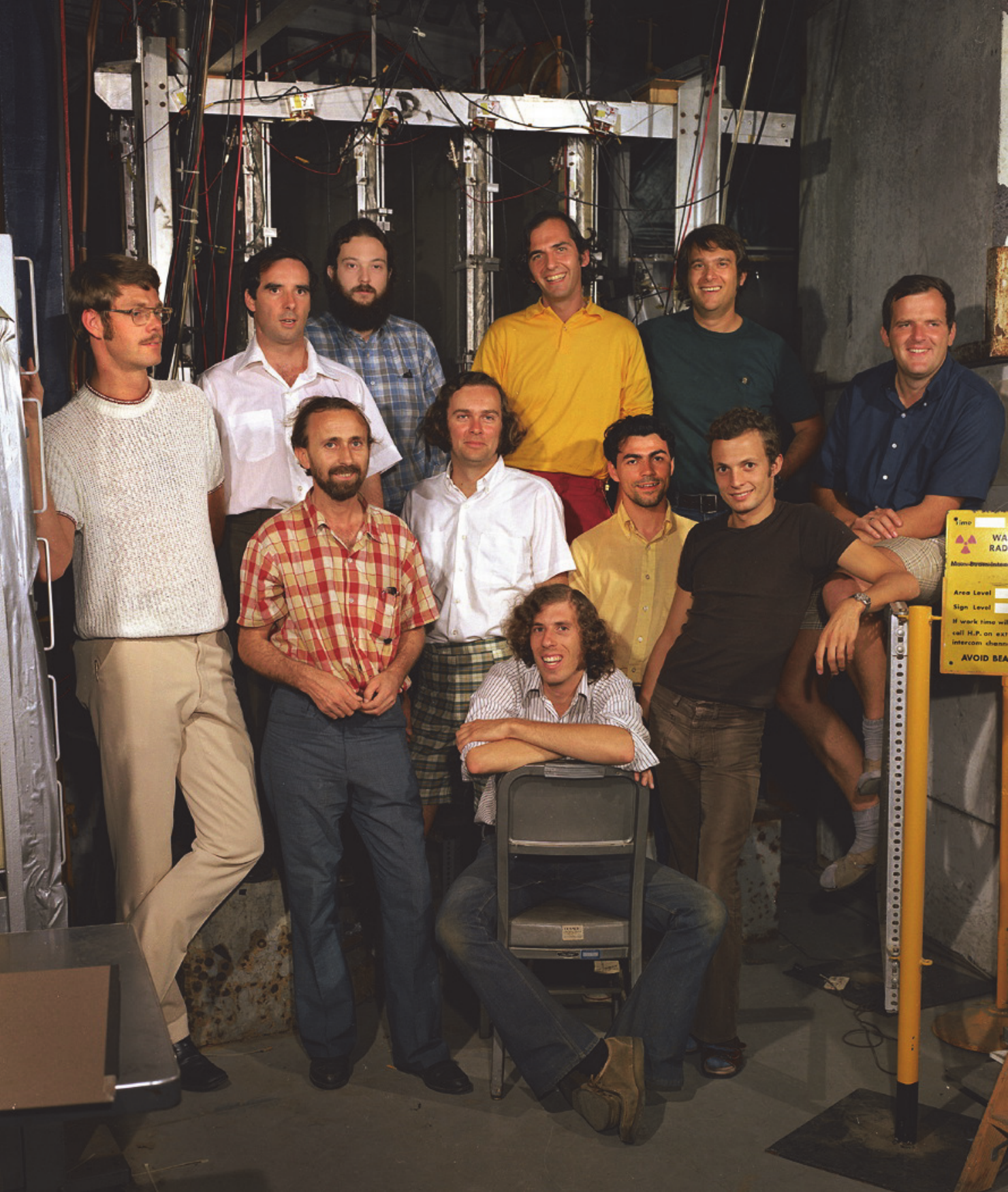}
\end{center}
\caption[]{Several Members of $\mu$-p II experiment, c. 1972. (Top Row-Left to Right) Tom Kirk, Mike Murtagh, Morgan May, MJT, Peter Limon, Hans Jostlein; (seated) Howard Gittleson. }
\label{fig:mu-p-II}\vspace*{-0.12in}
\end{figure}

	The period 1968--1971 was also the time of the startup of FERMILAB and I had joined 3 proposals: E70 with Leon to look for the $W^{\pm}$ in p+p collisions via single (and di-)leptons; E87 with Wonyong Lee to look for $W^{\pm}$ and heavy leptons, beyond the $e$ and $\mu$, in photoproduction (although I knew from my CEA experience that $e^+ + e^-$ collisions were better if they had sufficient c.m. energy); and E98, muon scattering, with the Harvard group. 
	In the meantime, I collaborated with Leon on another muon experiment at the AGS ($\mu$-p II, Fig.~\ref{fig:mu-p-II}) 
at which we made the first measurement of $\mu$-$A$ DIS thinking that if there were really point-like partons~\cite{BjorkenPaschos,Feynman1972} inside a proton the DIS cross section in $\mu$-$A$ for a nucleus $A$ should be $A$ times larger than for $\mu$-p (Morgan May's thesis)---we found $A^{1.00}$ for $x>0.1$, with some shadowing $A^{0.963\pm 0.006}$ for $x<0.1$, where $x$ is the fraction of the proton momentum carried by the struck parton~\cite{MMay}. With two Harvard graduate students, I also looked for excited muons (Howard Gittleson's thesis~\cite{HG}); and recoil hadrons from DIS (Mike Murtagh's thesis~\cite{MJM}) with the secret hope that the hadrons would show evidence of Bjorken and Feynman's point-like partons (they didn't). Mike later became chair of the BNL Physics department. 

Also, at this time I moved to the Rockefeller University where Rod Cool, who had been the associate laboratory director (ALD) for high-energy physics at BNL and collaborated on my thesis experiment, had moved and had hired me into his new group. Rod together with Leon Lederman had just proposed a $W^{\pm}$ search for $e^{\pm}$ produced at 90 degrees to the colliding beams at the new CERN Intersecting Storage Rings (ISR), a p+p collider with c.m. energy range $\sqrt{s}=22-62.4$ GeV. 
\section{Rockefeller University 1971-1980}
    The ``$W$'' experiment at the ISR got results in 1972, but, instead of finding an excess of high $p_T$ (transverse momentum) leptons perpendicular to the beam, found an excess of high $p_T$ pizeroes. The exponential $p_T$ spectrum observed in cosmic rays and paramaterized as $e^{-6p_T}$ by Cocconi~\cite{BBK} broke off and became a much flatter power-law for $p_T\geq 2$ GeV/c, thus greatly increasing the background for the $W^{\pm}\rightarrow e^{\pm}+X$ search. The high $p_T$ $\pi^0$'s proved that the partons from DIS interacted strongly with each other which couldn't be derived from the purely electromagnetic electron-proton (actually electron-quark) scattering. [This is the same idea as the search in muon-trident production for a muon-muon force which would not be evident in muon-proton scattering.]   The high $p_T$ announcement was the hit of the 1972 International Conference on High Energy Physics (ICHEP) in Chicago~\cite{FWB} and was clearly indicative of exciting new physics, so I resigned from  all my FERMILAB experiments and moved to Geneva to be with the Rockefeller group and continue the famous CERN-Columbia-Rockefeller (CCR) series of experiments. 
    
    We made several other discoveries in this series including direct single $e^{\pm}$ (the first evidence for charm ($c$) quarks~\cite{GIM}, but we didn't know it at the time)~\cite{CCRS}, direct single-$\gamma$~\cite{CCOR-gam1} and experimental proof using same and away-side two-particle correlations that high $p_T$ particles in p$+$p collisions are produced from states with two roughly back-to-back jets which are the result of scattering of the constituents of the nucleon as described by Quantum Chromodynamics (QCD), the theory of the strong interactions, which was developed during the course of these measurements (e.g. see~\cite{egMJTPrague}).  Interestingly, but not accidentally~\cite{MJT-Charm?}, these are the same techniques that we proposed and used 20 years later in the PHENIX experiment at the Relativistic Heavy Ion Collider (RHIC) at BNL (p+p to Au+Au collider at nucleon-nucleon c.m. energy $\sqrt{s_{NN}}=20-200$ GeV) in the initial search for (and discovery of) the Quark Gluon Plasma (QGP)~\cite{PXWP} since the huge multiplicity in Au+Au collisions gives an enormous combinatoric background which is minimal for single particle measurements and not too bad for two-particle measurements. 
    
    At the ISR, we also just missed discovering the J/$\Psi$, discovered by Sam Ting at BNL~\cite{Ting1974}, Burt Richter at SLAC~\cite{Richter1974},  and the $\Upsilon$, discovered by Lederman at FERMILAB~\cite{Lederman1977}, which are  quarkonia---bound states of $c-\bar{c}$ and $b-\bar{b}$ quarks. Note that we were beat out by lower energy machines with much higher luminosities even though the ISR had a respectable ${\cal{L}}=5\times 10^{31}$ cm$^{-2}$ s$^{-1}$ (DC) and could run as high as $1.4\times 10^{32}$ with the superconducting low $\beta$ quads (which had been assigned to another experiment).  
    
    It was during this same period that ISABELLE and the CERN SpS $\bar{\rm p}+{\rm p}$ collider were being proposed and built. Several of my European CCR colleagues (Luigi Di Lella and Marcel Banner) started the UA2 experiment at the SpS collider; but there were problems at the Rockefeller University, where the President Fred Seitz, a physicist, had attempted to eliminate the Physics and Philosophy departments, so I decided that ISABELLE would be a better bet and took the opportunity to return to BNL in 1980 to help sort out the problems they were having with the superconducting  magnets.  

\section{Brookhaven National Laboratory (BNL)-Superconducting Magnet Division 1980-82}
    I spent two years in the magnet division making and understanding superconducting accelerator magnets for ISABELLE, which was in a state of R\&D, i.e. no longer in construction, because the first 17 dipole magnets from industrial prodution failed to reach the design goal of 5 Tesla magnetic field although they all reached  $\approx 4$ Tesla after many quenches. There was fierce competition from FERMILAB, 
    where Leon Lederman had become Director in late 1978. FERMILAB also made superconducting magnets for their ``doubler'' (which later became the FERMILAB Tevatron $\bar{\rm p}+{\rm p}$ collider) that were made from the much more robust ``Rutherford cable''.  These magnets also had problems but they were cured in late 1980~\cite{Hoddeson1987}. Bob Palmer from the BNL physics department figured out how to use two layers of FERMILAB cable, plus Cu wedges for field shaping, to make ISABELLE magnets that were much superior to FERMILAB's in both field quality and the ability to reach the maximum current in the (Type-II) NbTi superconductor without quenching; and we switched over to making these magnets in September 1981. This design became the basis for all post-Tevatron superconducting hadron accelerators including HERA, SSC, RHIC and LHC. Again, Bob Crease's articles on ``The ISABELLE saga''~\cite{Crease-saga}'' recall this exciting and intense period in great detail.     
    
\section{BNL Physics Department-1982-present} 
\subsection{Snowmass 1982, ICHEP 1982}    
With the success of the Palmer magnet, I got transferred to the Physics department in 1982 where two miracles happened. While at the magnet division, I had requested to attend the 1982 ICHEP (in  Paris, France) and was submitted by BNL as an alternate. The first miracle is that Mark Sakitt, who was one of the official BNL delegates, decided not to go. I understood from Kit D'Ambrosio (who was Rod Cool's secretary when he was ALD and was still the ALD secretary in 1982) that the ALD (Nick Samios) originally didn't want me to be the substitute for Mark but she insisted that since BNL had told the French that    I was the alternate, BNL had to send me to avoid an international incident. The second miracle was that Dave Levinthal (at Columbia) and Steve Pordes (at FERMILAB) working on the CCOR data in the two years that I was making magnets had figured out why the pizero-pair data, which I had suggested be analyzed in the same variables (pair-$p_T$, pair-invariant mass, pair-rapidity and angle in the pair c.m. system) that were used in lepton-pair analyses, had too steep a $\cos\theta^*$ dependence. Because this process corresponded to quark-quark elastic scattering in the newfangled QCD, the \mbox{$Q^2=-\hat{t}\approx m^2_{\pi^0 \pi^0} (1-\cos\theta^*)/2$} is reduced for angles forward of 90$^\circ$ ($\cos\theta^*=0$). This causes the running coupling constant $\alpha_s(Q^2)$ to increase at more forward angles, which steepens the angular distribution. The incredible thing about this data is that the way we had constructed the $\pi^0$-$\pi^0$ $\cos\theta^*$ distribution, it could be directly compared to the parton-parton constituent  level scattering cross-section in QCD without need for a Monte Carlo (although of course we did the Monte Carlo and showed that this was true). 

   I showed these results at the famous Snowmass conference of 1982~\cite{ProcSN82} that was organized by the Division of Particles and Fields (DPF) of the American Physical Society (APS) to consider the future facilities for particle physics in the U.S. which included proposals for $e^+ e^-$, $e+$p and hadron-hadron colliders and was attended by 150 physicists from all these fields. There was a matrix of physics categories (e.g. specific standard model tests at the different machines) and I headed the QCD-dynamics subgroup, ``Isolating and testing the Elementary QCD subprocesses'', of the standard model group, with a view, for instance, of testing the recently published formulas for hard scattering of the underlying partonic subprocesses in pQCD, e.g. $q+q\rightarrow q+q$, $g+q\rightarrow \gamma+q$, in analogy to precision tests of QED~\cite{MJT-SN82}. However, although jets of hadrons had been well established in the reaction $e^+ +e^-\rightarrow q +\bar{q}\rightarrow {\rm hadrons}$~\cite{Hanson1975}, there were doubts about the existence of jets in hadron-hadron collisions in the U. S. HEP community gathered at Snowmass.     These doubts were caused by an erroneous claim for jets in p+p collisions at FERMILAB in 1977  that was contradicted by a result by NA5 (CERN), presented at the 1980 ICHEP\cite{Pretzl1980,DeMarzo1982} (e.g. see a review in~\cite{MJTIJMPA}). The NA5 result showed that there were no jets in large aperture full azimuth calorimeters triggered on the sum of transverse energy ($E_T$), or in their words, ``the large $E_T$ observed is the result of a large number of particles with a rather small transverse momentum''. Hence, I could not convince the lepton physicists and even had difficulty convincing the hadron physicists of the value of my plot or even of the utility of QCD tests in hadron-hadron collisions since everybody was much more interested in the Electro-Weak side of the standard model, namely discovery of the $W^{\pm}$. My main role in the strong BNL effort on this subject (which had been the principal goal for HEP and certainly ISABELLE at this time) was in understanding the background from hard-scattered hadrons. 
   
\begin{figure}[!htb]
\begin{center}
\includegraphics[width=0.95\linewidth]{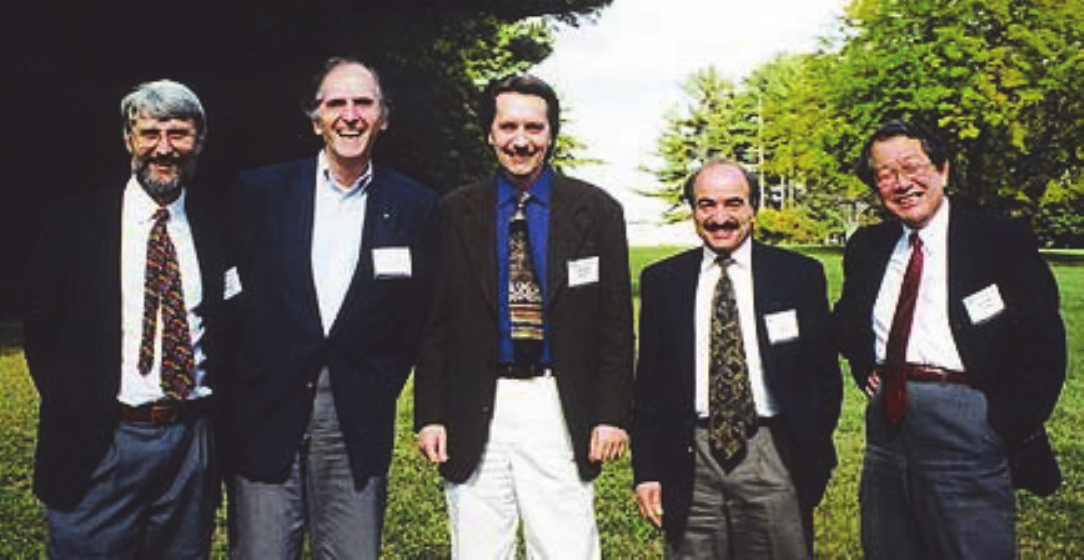}
\end{center}\vspace*{-0.12in}
\caption[]{Some founding members of RSC-Bunce, MJT, Roser, Makdisi, Ozaki }
\label{fig:RSC}\vspace*{-0.12in}
\end{figure}

   As I worked on the $W^{\pm}$, I became more interested in the parity-violating aspect, which led to my interest in later founding the RHIC spin collaboration (RSC) along with Gerry Bunce, Yousef Makdisi, Thomas Roser, with encouragement from Satoshi Ozaki (Fig.~\ref{fig:RSC}), Larry Trueman, Sam Aronson, Tom Kirk and many others. 
My interest in spin was originally piqued by my having attended the 1980  Symposium on High Energy Physics with Polarized Beams and Polarized Targets in Lausanne as (of all things) a representative of FERMILAB since I had designed a coherently hardened photon beam (which was also polarized) to enhance the high energy tail of the brems-spectrum for E87~\cite{MJT-Spin80}. Interestingly, in reviewing my foreign travel report from 1980, I found that I had made the recommendation:``A serious study of operating ISABELLE with polarized protons should be performed including the Accelerator side and the Physics side, in particular polarization effects in $W^{\pm}$ and $Z^0$ production'', perhaps because Ernie Courant presented a talk with the title ``Polarized Protons for ISABELLE''~\cite{Courant1980} at this meeting.  

\begin{figure}[!htb]
\begin{center}
\begin{tabular}{cc}
\includegraphics[width=0.38\linewidth]{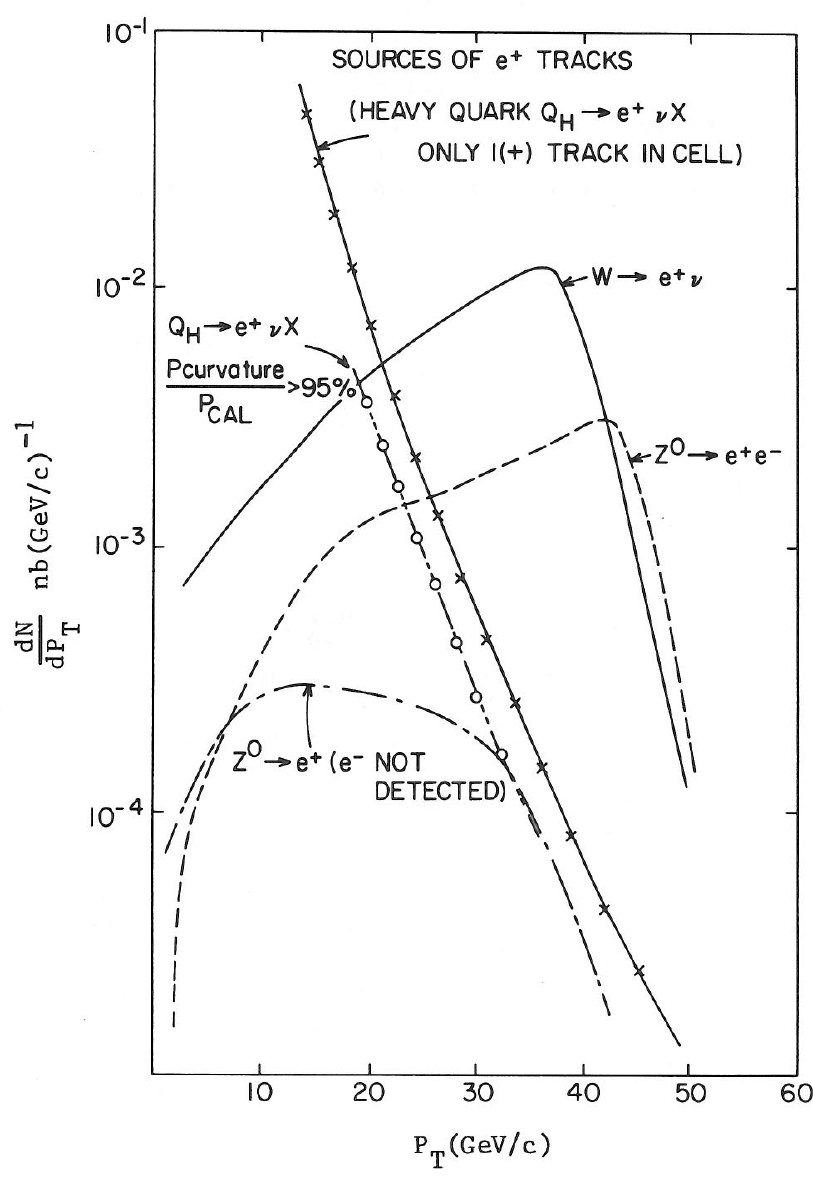}&
\hspace*{0.02\linewidth}\includegraphics[width=0.47\linewidth]{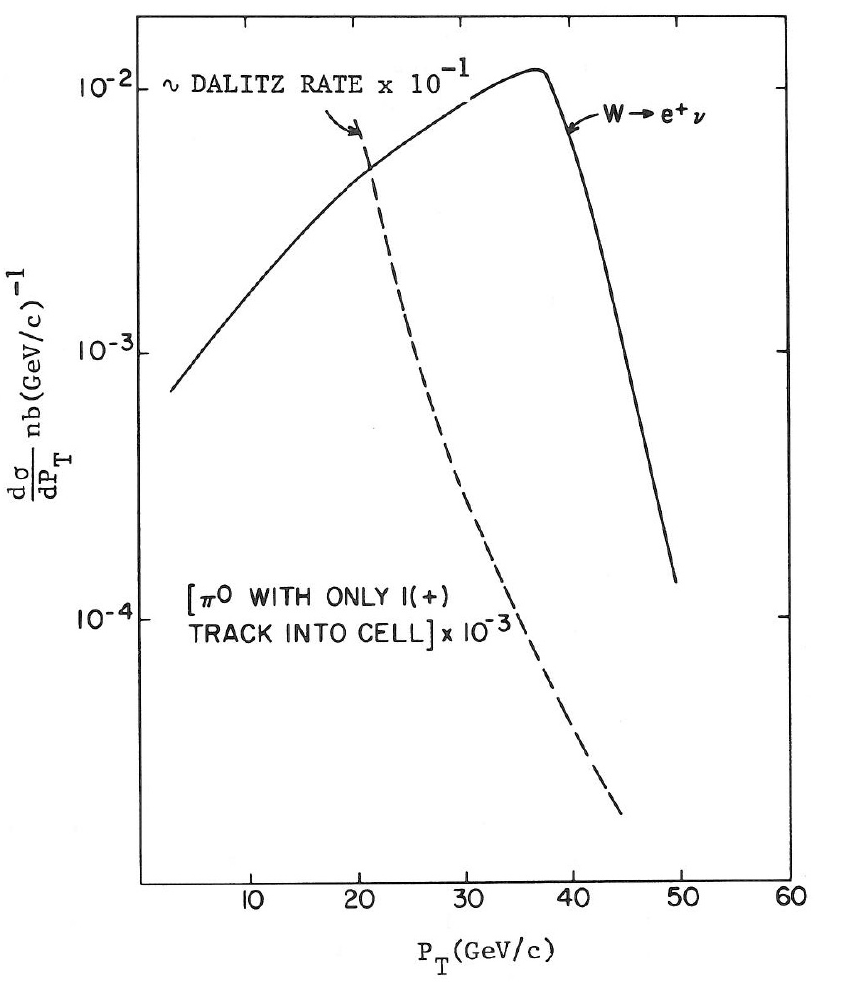}
\end{tabular}
\end{center}\vspace*{-0.25in}
\caption[]{BNL $W^+\rightarrow e^+ +X$ plots from Snowmass 1982~\cite{BNLWSN82}  with (left) expected heavy quark background (right) expected hadron backgrounds.  }
\label{fig:Snowmass82}\vspace*{-0.12in}
\end{figure}

	One of the BNL contributions to the Snowmass proceedings contained the following plots (Fig.~\ref{fig:Snowmass82})~\cite{BNLWSN82} of the $W\rightarrow e+X$  Jacobian peak at mid-rapidity at $\sqrt{s}=800$ GeV together with the backgrounds from heavy quarks, photon conversions and $\pi^0$-charged hadron overlaps, including the estimated rejection factors, with the conclusion that, ``it appears straightforward to reduce the backgrounds of single positrons so that the $W^+$ contribution to the high $p_T$ single positron spectrum is dominant above $p_T$ of 20 GeV/c. We have not included transverse momentum balance in the above discussion.'' Since we actually knew all the cross sections in 1982 and had a really good idea of the $W$ mass, these  calculations should be still valid today for RHIC. 	
Larry Trueman and Gerry Bunce also introduced a discussion of polarized proton collisions in the same article, their point being that one could pick out the hadronic decay $W^{\pm}\rightarrow$ di-jet signal from the large hadronic background by using a parity violating subtraction which would eliminate the background~\cite{Paige1979}. Another interesting article at this conference~\cite{Abolins1982} was a conjecture that the several generations of quarks and leptons could be composites of more fundamental constituents, with a scale of compositness $\Lambda$ greater than 100 GeV. The intriguing feature of such composite models is that the interactions generally violate parity because the scale is larger than the weak interaction scale, the mass of the $W^{\pm}$. This would result in a deviation from QCD at large $p_T$. It is difficult to prove that a small deviation is really due to something new, but if it was found that the difference from QCD were parity violating, this would be a clear signature of new physics.
     
\begin{figure}[!ht]
\begin{center}
\begin{tabular}{cc}
\hspace*{-0.1in}\includegraphics[width=0.75\linewidth]{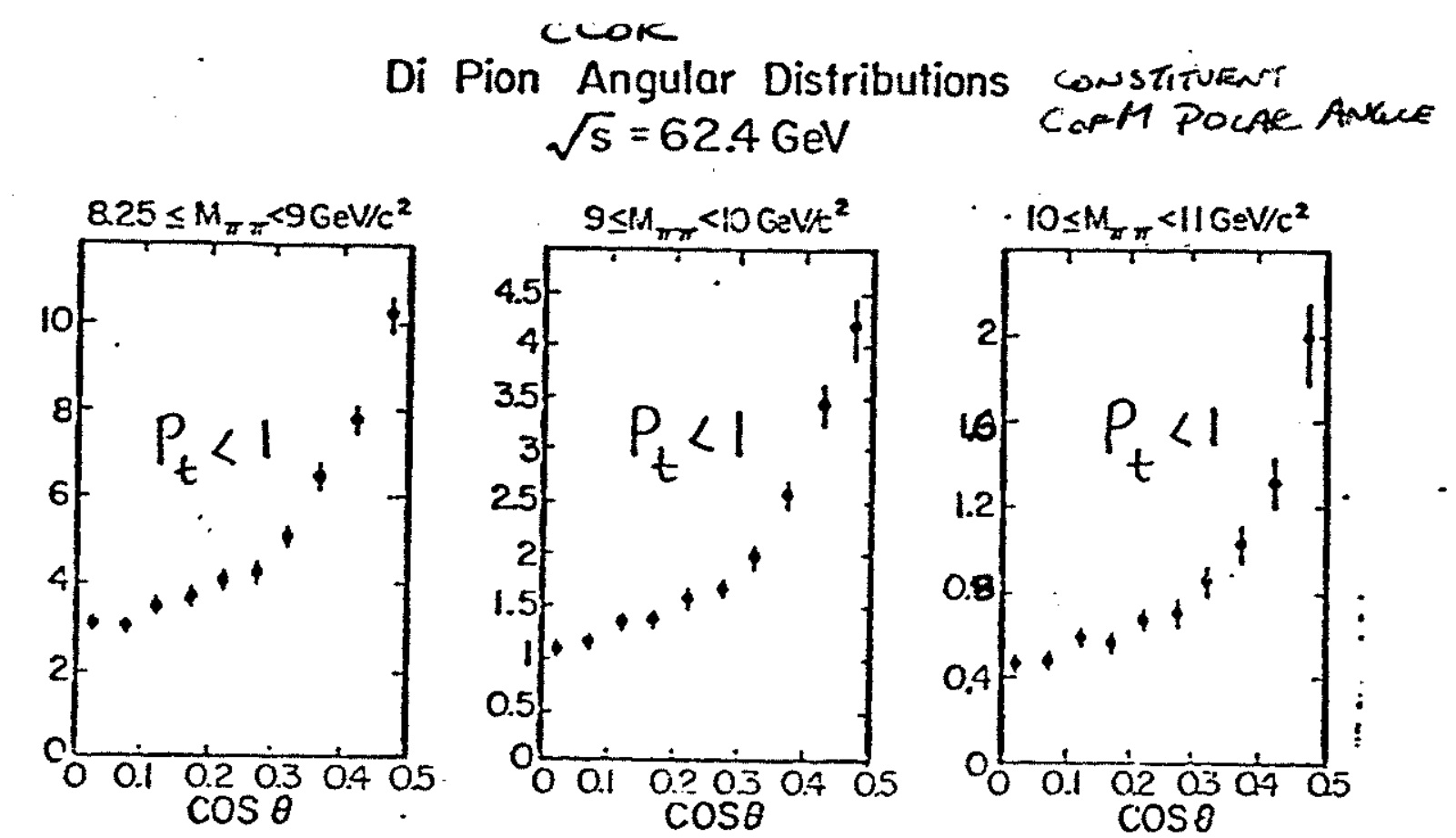} &
\hspace*{-0.35in}\raisebox{-1.0pc}{\includegraphics[width=0.288\linewidth]{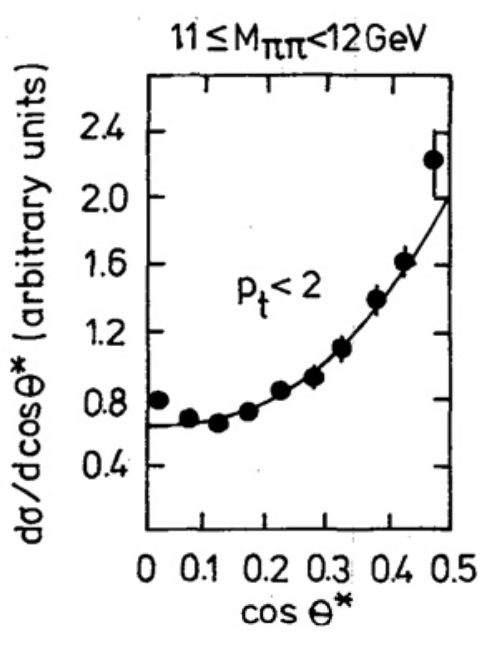}}
\end{tabular}
\end{center}
\caption[]
{a) (left 3 panels) CCOR measurement~\cite{CCOR82NPB,MJTParis82} of polar angular distributions of $\pi^0$ pairs with net $p_T < 1$ GeV/c at mid-rapidity in p+p collisions with $\sqrt{s}=62.4$ GeV for 3 different values of $\pi\pi$ invariant mass $M_{\pi \pi}$. b) (rightmost panel) CCOR measurement with QCD predictions for the elastic scattering of $q$-$q$ with $\alpha_s(Q^2)$ evolution~\cite{WolfParis82}.    
\label{fig:Paris82} }
\end{figure}

	Back at my new office in the Physics Department, after Snowmass, before going to Paris, I calculated the formulas for the QCD parton-parton scattering predictions which I had spent the last few weeks studying and plotted them onto our $\pi^0$-pair data---the agreement with $q+q$ scattering was unbelievably beautiful once the running of $\alpha_s(Q^2)$ had been taken into account (Fig.~\ref{fig:Paris82}, Fig.~\ref{fig:quarkstuff}a).

When I got to Paris, I bumped into Carlo Rubbia who was scheduled to present  the conference highlights talk. I asked him to show my plot but he said something like: ``Get lost, I have much better things to show''. I was pretty disappointed and I went into the room where all the submitted contributions were on display and I met Knut Hansen (an ISR physicist from Niels Bohr Institute) who was organizing the Hard Hadron Physics parallel session. He said that CCOR had submitted a nice paper on $\pi^0$-pairs~\cite{CCOR82NPB}, would I like to give a talk on it? [A third miracle, I guess.] Well it turned out that I gave my talk~\cite{MJTParis82} after the session at which UA2 had presented their famous lego plot of a jet,  which convinced everybody that jets existed, so I got great exposure as evidenced by the fact that Carlo [who had been completely blindsided by the UA2 jet] came to me and asked for the plot to show in his talk as did G\"unter Wolf who was the rapporteur on Jet Production and Fragmentation (Fig.~\ref{fig:Paris82}b). As they say, the rest of the story on jets and QCD is history; but it is unfortunate that the UA2 result was not available at the Snowmass meeting.    

   One of the sidelights to Snowmass '82 was the contribution to the proceedings  by Huson, Lederman and Schwitters~\cite{Primer} (not presented at the meeting) which claimed that one couldn't do an experiment at $10^{33}$ cm$^{-2}$ sec$^{-1}$ unless ``the apparatus was shielded from the collision region by massive quantity of steel.'' This, in my opinion, was a blatant attempt to discredit the 800 GeV, $10^{33}$ cm$^{-2}$ s$^{-1}$ ISABELLE project compared to the proposed 2000 GeV $10^{30}$ cm$^{-2}$ s$^{-1}$  Tevatron $\bar{\rm p}+{\rm p}$ collider. As somebody who had been beaten to both the $J/\Psi$ and the $\Upsilon$ by lower c.m. energy, much higher luminosity (fixed target) machines, I seemed to be the only person at BNL who strongly objected to this claim (in writing). Another thing that galled me was that Leon, who had originally proposed the $10^{33}$ cm$^{-2}$ s$^{-1}$ luminosity for ISABELLE 10 years earlier, now seemed to be saying that it wouldn't work.  One good side of this debate (which I seem to have won in a classic example of a pyrrhic victory because the LHC is designed for $10^{34}$ cm$^2$ s$^{-1}$ with large open geometry detectors) is that the issue of pile-up became important and in a follow-up 1983 DPF-APS workshop Randy Johnson~\cite{Randy83} presented an elegant analytical solution for calculating pile-up right about the same time that we got $\alpha+\alpha$ collisions in the ISR. This led me to write a CCOR internal memo (R-110:171)  with the title, ``How Leon Lederman drove me into convolutions and why this helped me to understand $\alpha+\alpha$ collisions'' which resulted in much fun with Gamma and Negative Binomial distributions working with Chellis Chasman and John Olness in understanding the $\alpha+\alpha$ data and later in heavy ion (A+A) collisions at the AGS and at RHIC---physics programs that were beyond our thinking at that point but miraculously followed from the dramatic events as 1983 progressed.  One interesting point about my entry into heavy ion collisions with the $\alpha+\alpha$ run at the ISR (proposed by Martin Faessler and others~\cite{ISRC}) is that we were faced with either shutting off CCOR for a week and taking a short rest or running for that week and getting new physics and dozens of new papers. That's really no choice for a physicist---we ran the detector and took data; and exactly the same thing happened at the LHC.   
\subsection{1983--1988 ISABELLE $\rightarrow$ CBA $\rightarrow$ RHIC}   
   The first dramatic event in 1983 was the announcement by Carlo Rubbia at an APS meeting in New York City on January 26, and the previous week at CERN, that the UA1 experiment at the CERN SpS collider~\cite{UA1W} had discovered the $W^{\pm}$ boson, followed shortly by confirmation from UA2~\cite{UA2W}. This was followed by meetings of the High Energy Physics Advisory Panel and subpanels that resulted in the cancellation of ISABELLE (whose  name had been changed to the Colliding Beam Accelerator, CBA, after the Snowmass '82 meeting) on July 11, 1983. On this very same day another miracle occurred: the Nuclear Science Advisory Committee (NSAC) which was having a workshop to recommend their next major construction project seized the opportunity and proposed to build a colliding beam heavy ion accelerator in the CBA tunnel. Within a day or two, Nick Samios, the BNL director, with the influence of T.~D.~Lee, appointed a task force to adapt the ISABELLE design into a Relativistic Heavy Ion Collider, which the task force named RHIC. Once again, the rest is history (which is also well described by Bob Crease~\cite{Crease2008}), with the proposal for RHIC formally submitted in 1984, approved for construction in 1990 with the first collisions in 2000.  
   
\begin{figure}[!ht]
\begin{center}
\begin{tabular}{cc}
\hspace*{-0.1in}\includegraphics[width=0.40\linewidth]{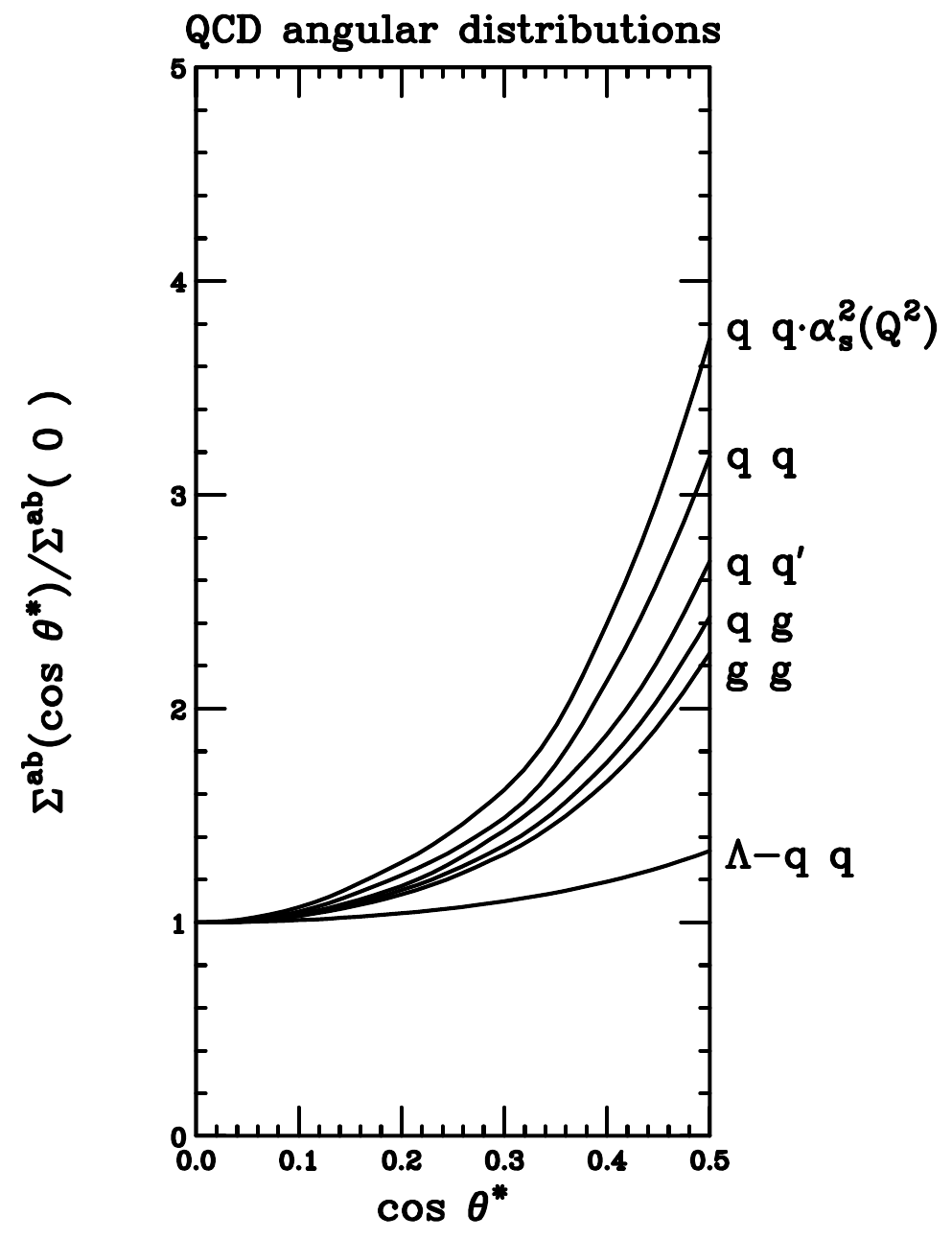} &
\hspace*{-0.15in}\raisebox{-1pc}{\includegraphics[width=0.55\linewidth]{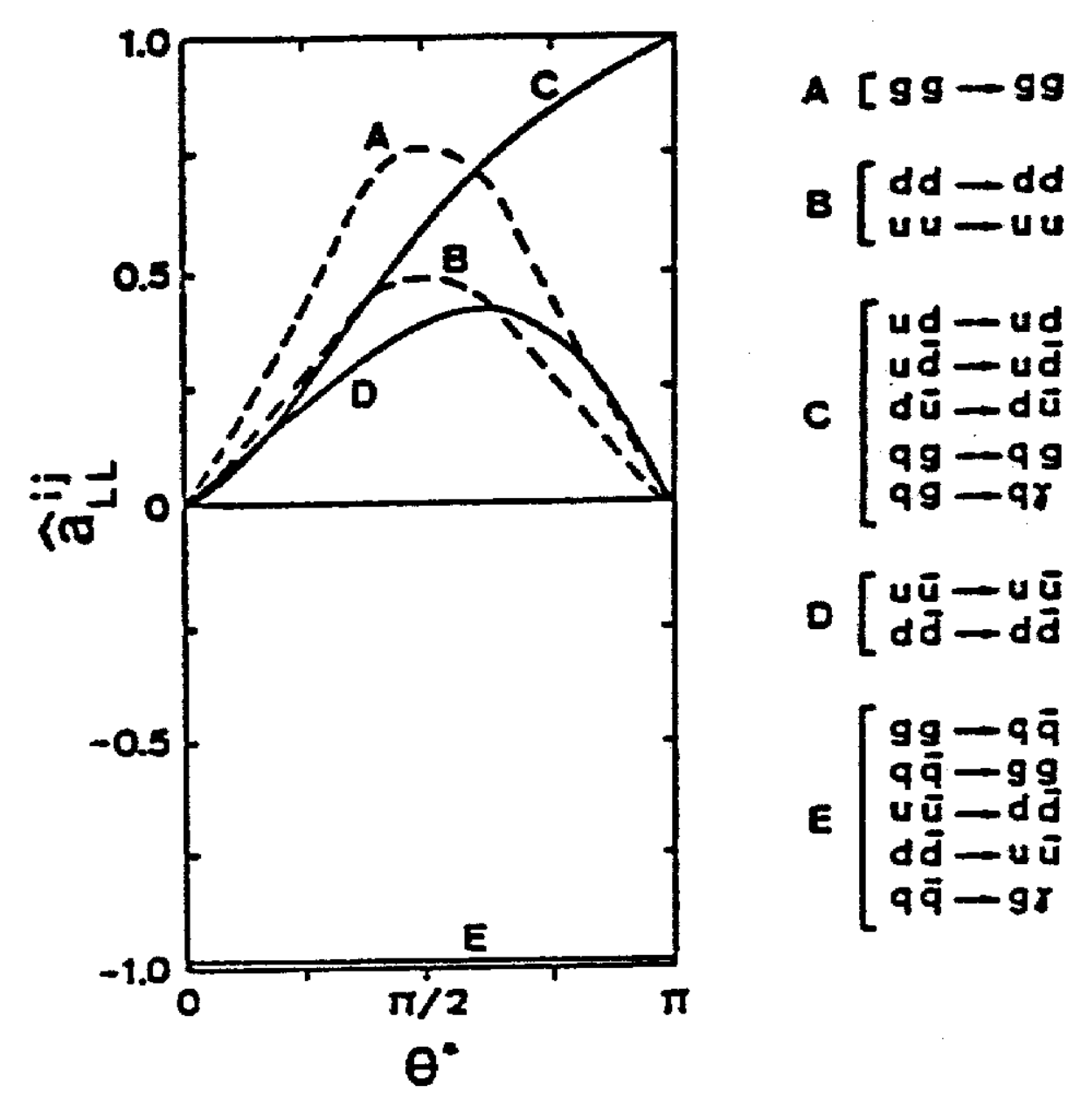}}
\end{tabular}
\end{center}
\caption[]
{a) (left) 
 QCD predictions for the elastic scattering angular distribution of $g+g$, $q+g$, $q+q'$, $q+q$, and $q+q$ with $\alpha_s(Q^2)$ evolution. The effect of the interference of a possible contact term with $\Lambda$ on the $q+q$ scattering distribution~\cite{Abolins1982} is also shown. b) Longitudinal two-spin asymmetry $\hat{a}_{\rm LL}$ as a function of scattering angle for the subprocesses indicated~\cite{Sivers1979}.    
\label{fig:quarkstuff} }
\end{figure}

   Once the $W$ was discovered, I started thinking about ISABELLE/CBA in terms of hadron physics, especially polarized $p+p$ collisions as a still unexplored domain where new tools (parity violation) could lead to the discovery of new, unexpected phenomena such as the parity-violating quark substructure that was discussed at Snowmass. Also QCD is a gauge theory of the srong interactions in which helicity plays as fundamental role as charge and there are QCD predictions for the two-spin longitudinal asymmetry of charged partons~\cite{Sivers1979}, equally as important to test as the parton-parton elastic scattering cross sections (Fig.~\ref{fig:quarkstuff}).
 
    There was already a polarized proton beam collaboration at the BNL-AGS, led by Alan Krisch; and in May 1983, I presented a talk at their collaboration meeting with the title ``Measuring and using Polarized Protons at CBA'' based on work of Frank Paige and myself~\cite{FPMJT}. In November 1983, several months after CBA became RHIC, I put all my ideas into a long memo with the innocuous title ``Auxiliary facilities for a Relativistic Heavy Ion Collider''~\cite{MJTAux} which I distributed to key decision makers; and I decided to join the Heavy Ion Program at the AGS, which was the first step leading to RHIC. Sure enough, during the 1983-84 design period of RHIC the issue of whether to allow $p+p$ collisions was raised by both FERMILAB who was afraid of competition and some influential Nuclear Physicists who thought that this was just an excuse to ``ressurect the CBA''~\cite{Baym2002}.  Fortunately, it turned out that neither FERMILAB nor the Nuclear Physicists objected to the polarized $p+p$ collider option.

	The termination of ISABELLE/CBA led the High Energy Physics community to start taking seriously the idea of a $\sqrt{s}=20-40$ TeV Superconducting Super Collider (SSC) which had started as Bob Wilson's ``desertron'' at Snowmass '82~\cite{LML82}, so I still kept on thinking about the $W$. For an SSC fixed target workshop in 1984~\cite{MJTSSCF}, I designed a 9.48 TeV coherently hardened photon beam for the purpose of using single $W^+$ photoproduction, $\gamma+p\rightarrow W^+ +n$, to measure the tri-linear $\gamma W^+ W^-$ coupling. The only good to come of this is that I realized that the existing calculations at that time had made the same mistake as in the original calculation of $p+p\rightarrow d +W^+$: they left out the nucleon form factors. The calculations were corrected~\cite{Dicus}  but the measurement was eventually done at HERA, with $e+p\rightarrow e + W + X$~\cite{SouthTurcato2016}, and also with $e^+ +e^+\rightarrow W^+ + W^-$ at LEPII~\cite{DELPHI2010}.  

\subsection{RHIC Spin Collaboration $\rightarrow$ RIKEN-BNL Research Center, 1988--1995$\rightarrow$}	
    In 1988, Peter Bond, chairman of the Physics department, organized 
a committee to study the ``Future of 
High Energy Physics at BNL". Being a member of this committee, I took 
the charge literally and pushed for a polarized proton facility at RHIC, 
which was accepted by the committee which reported in early 1989 and 
recommended a ``Task force on Polarized Protons at RHIC...to include 
accelerator physicists to come up with a realistic scheme for manipulating 
and maintaining the polarization. It should include theorists... One would 
like to lay out the potential physics program and determine what kind of 
a detector could serve it best." 

Pursuant to this recommendation, Sam 
Aronson---then deputy chairman of the Physics Department---convened a group 
to discuss this subject and reported to Larry Trueman, the ALD, in March 1989, requesting his input before setting up such a task force. This approval was given and a task 
force was set up in April 1989, with Gerry Bunce and 
Mike Tannenbaum as the co-leaders. The subject of Polarized Proton Physics 
at RHIC was also mentioned in the official BNL presentation to the HEPAP 
Subpanel on Future directions in High Energy Physics in January 1989. 

The task force was being 
organized in May 1989, when Satoshi Ozaki---who had left BNL in 1981 for Tsukuba, Japan to lead the construction of TRISTAN, a $\sqrt{s}=60$ GeV Electron-Positron Collider at the National Laboratory for High Energy Physics (KEK)---returned to BNL in 1989 as the RHIC Project Director. He immediately decreed that there should be nothing public  (i.e. outside BNL) 
on ``polarized protons at RHIC'' until the RHIC project was officially 
approved. This somewhat stifled the task force.    
However, in January 1990, two significant things happened: President Bush 
officially put RHIC into the DOE (U.S. Department of Energy) budget, and Shysh-Yuan (S. Y.) Lee and Ernie  Courant figured out how to inject polarized protons into RHIC with the existing  
injection line. 

The next great milestone occurred in late February, when 
the agenda for the HEPAP subpanel (of 1990) at BNL came out and I noticed 
that Satoshi Ozaki was making a presentation entitled 
``Proton Opportunities at RHIC''. I called this to Gerry's attention, and 
the next thing I knew (February 28, 1990 at 11:00) I was in Ozaki's office attending 
a meeting with Gerry, Satoshi and Tom Ludlam on unsupressing the polarized 
task force. This was successful and led to the  Polarized Collider Workshop held 
at Pennsylvania State University, University Park Campus, November 15-17th 
1990~\cite{PSU1990} which resulted in the formation of the RHIC Spin Collaboration (RSC--official start date is Jan 1, 1991) a group of experimental, theoretical and accelerator physicists, with a common interest in spin, to add polarized proton capability to RHIC. 

At the first Program Advisory Committee (PAC) to evaluate Letters of Intent (LOI) for experiments at RHIC, 
in August 1991, the RSC LOI was presented by Gerry Bunce along with the heavy ion proposals called OASIS, Dimuon, TALES/SPARHC, STAR and an AGS proposal E880, to study  ``The Effects of 
a Partial Siberian Snake on Polarization at the AGS''. The STAR experiment and E880 were approved and the new ALD Mel Schwartz\footnote{1988 Nobel Prize winner along with Leon Lederman and Jack Steinberger for the discovery of the muon-neutrino at BNL in AGS-28, the experiment that I turned down.} told the three other experiments to merge into an experiment (most like TALES/SPARHC) to study electrons and photons emerging from the Quark Gluon Plasma, which became the PHENIX experiment, with Sam Aronson as Project Director who appointed Shoji Nagamiya (then at Columbia) as spokesperson. Two other smaller experiments were also approved (see Refs.~\cite{NIM499,RATCUP} for details.) 

The RSC proposal was independent of the approved large experiments which both later decided that they wanted to participate. This led Mel Schwartz to make a general call for spin proposals for the PAC meeting in September 1992 at which a full proposal from the RSC for a program of Spin Physics using the RHIC polarized Collider was presented by Thomas Roser which included a general section---covering an overall view of the physics and a detailed conceptual design for the spin rotators, siberian snakes and polarimeters which would be necessary to operate RHIC with polarized protons---followed by specific spin proposals presented by both the STAR and PHENIX collaborations. The result was that ``The PAC was impressed with the work done but felt that approval would be premature.'' This was sufficient for Mel to insist that spin not be excluded in the design of RHIC. The following PAC meeting in February 1993 concluded that they were ``impressed with the compelling physics'' and requested  a formal proposal for ONE detector and ONE set of spin rotators.  After much convincing by RSC members, Mel Schwartz agreed that a second set of spin rotators is a bargain and accepted a two-rotator/two-experiment spin program ``if the  technical review of polarized protons at RHIC is successful''.
 
\begin{figure}[!htb] 
	\begin{center}
\begin{tabular}{cc}
\psfig{file=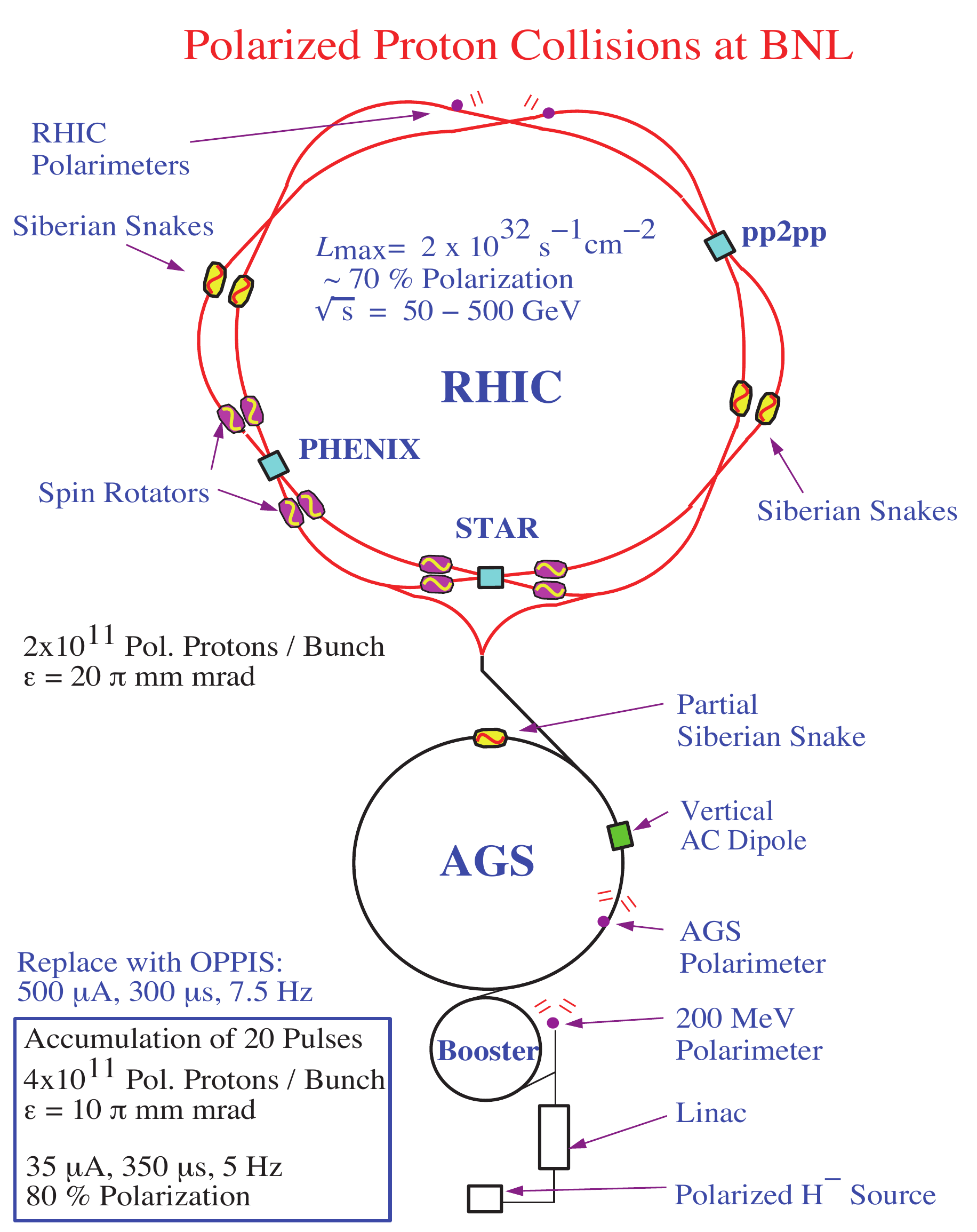,height=3.1in}
\hspace{0.1in}
\psfig{file=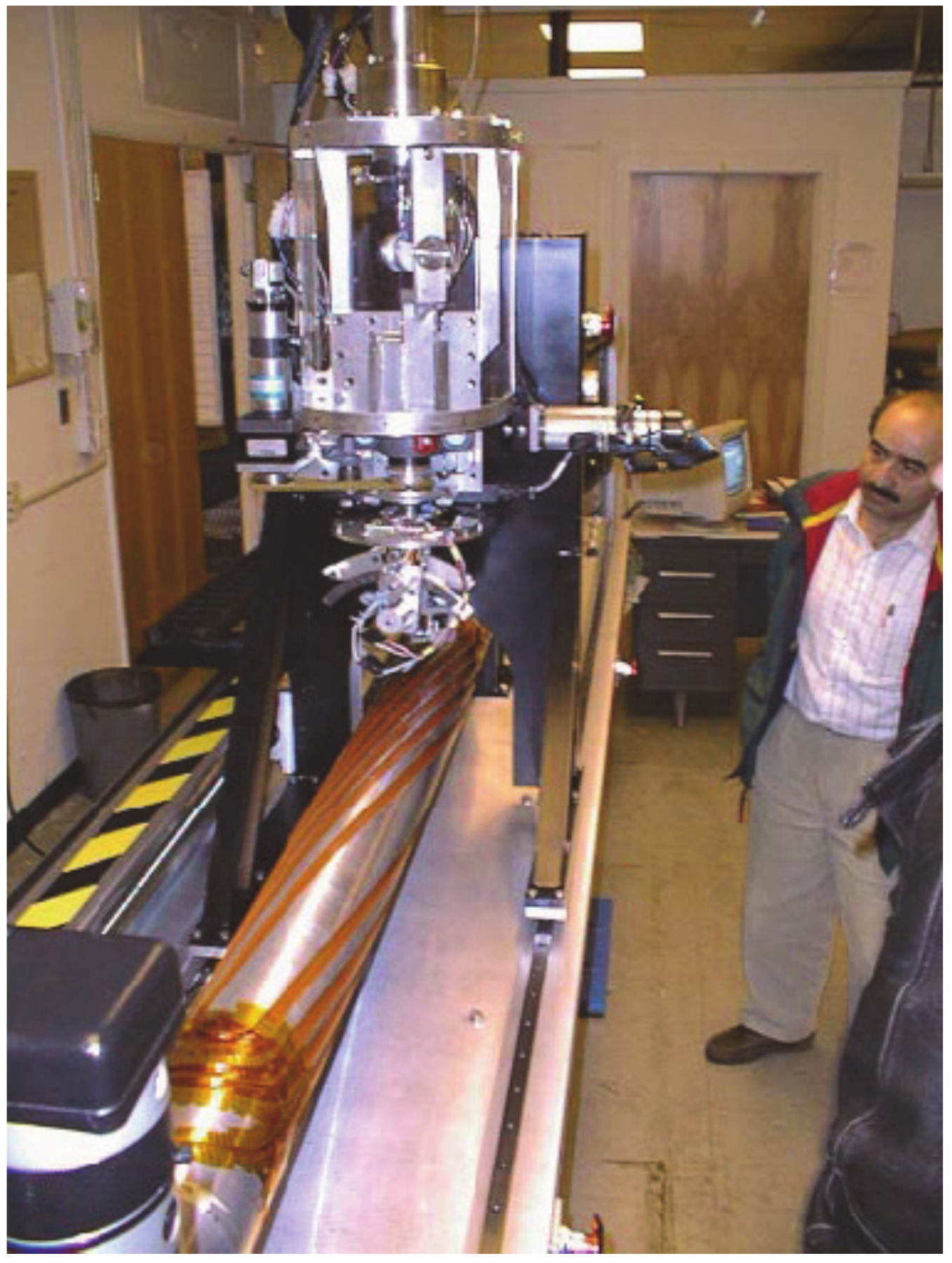,height=3.1in} 
\end{tabular}
\caption[]{ a) (Left) RHIC machine with polarized proton hardware highlighted~\cite{Alekseev2003}. b)(right) Yousef Makdisi watching siberian snake magnet being wound at BNL.   
\label{fig:RHICmachine}}
\end{center}
\end{figure}

The Technical Review in June 1993 concluded that ``The proposal has the flavor of the application of an ingenious technological invention 
(siberian snakes) to make possible exciting physics research (polarization 
physics) reminiscent of the application of stochastic cooling to obtain 
$\bar{p} p$ beams for $W$ and $Z$ in the CERN SPS.'' This was in hindsight a truly enthusiastic endorsement: the CERN project was awarded the Nobel Prize the next year. The other good news from Hideto En'yo, one of our Japanese colleagues who was a member of both PHENIX and the RSC, was that there was possibly new interest in RHIC Spin physics in Japan.  This eventually led RIKEN, Japan's largest and most comprehensive research organization for basic and applied science, to agree to fund the spin hardware in RHIC including \$10M for the siberian snakes\footnotemark\ and spin rotators and \$10M for a second muon arm for the PHENIX experiment in September 1995 (Fig.~\ref{fig:RHICmachine}). Interestingly, the same thing then happened as happened after the February 1993 BNL PAC meeting, namely, when Akito Arima, the President of RIKEN, was at BNL finalizing the arrangements, Gerry and I again got a call to Ozaki's office to answer the question of why RIKEN should provide spin rotators for the STAR experiment when they were only supporting PHENIX. We explained again that the approved project was for both experiments and he understood and agreed. A perhaps more important result of the collaboration with RIKEN was their establishment and funding of the RIKEN BNL Research Center in April 1997, located at BNL, dedicated to the study of strong interactions including spin physics, lattice QCD and RHIC physics by nurturing a new generation of young physicists, with founding Director T.~D.~Lee.

\footnotetext{A siberian snake is a magnet that flips the spin of the protons half way around the ring to preserve the polarization by cancelling imperfections. The spin is aligned vertically along the magnetic field in RHIC, perpendicular to the beam direction, and can be rotated to be longitudinal at the two interaction points.}
\subsection{We also learned some physics 1990 - present} 
\begin{figure}[!htb] 
	\begin{center}
\begin{tabular}{cc}
\psfig{file=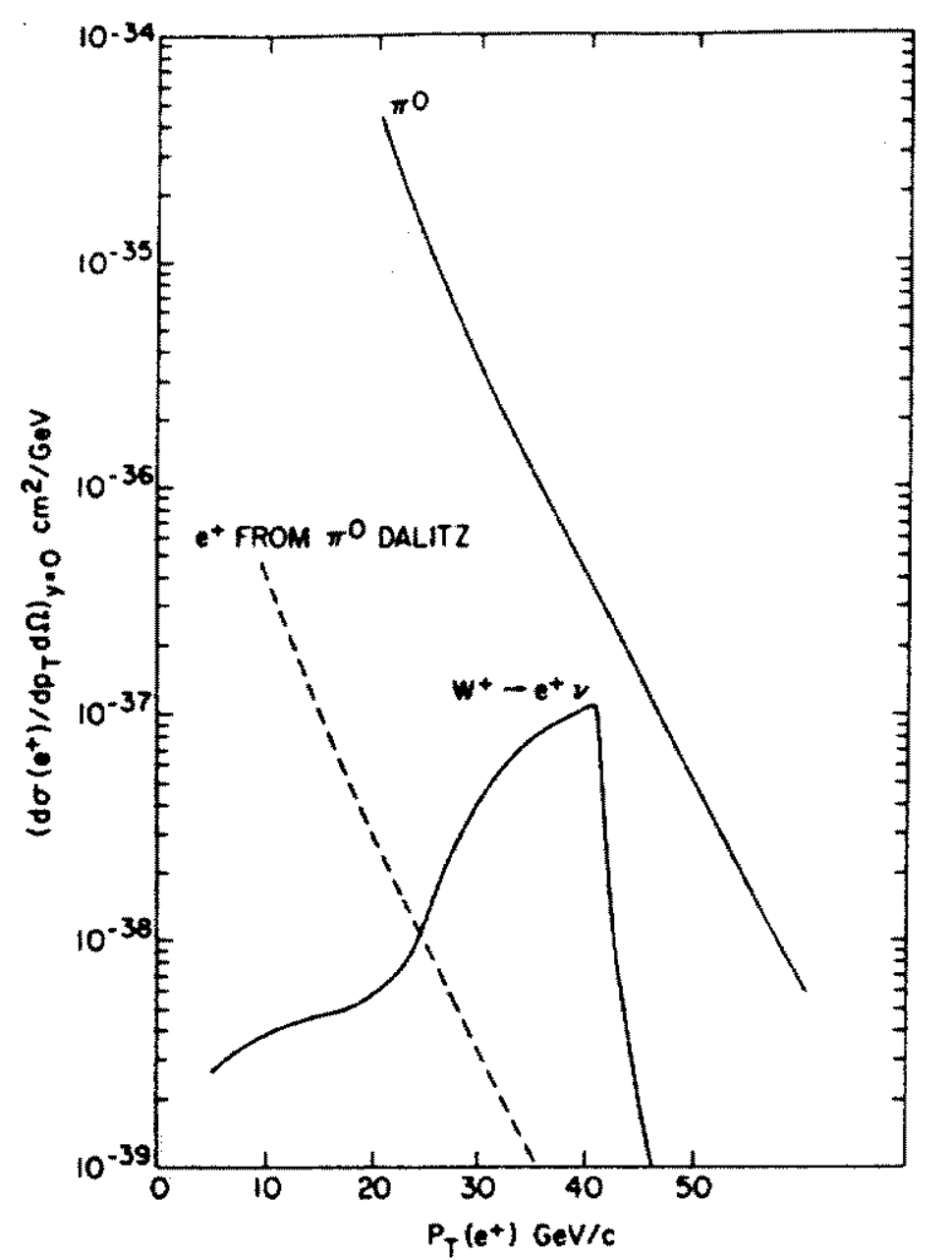,height=3.2in}
\hspace{0.1in}
\psfig{file=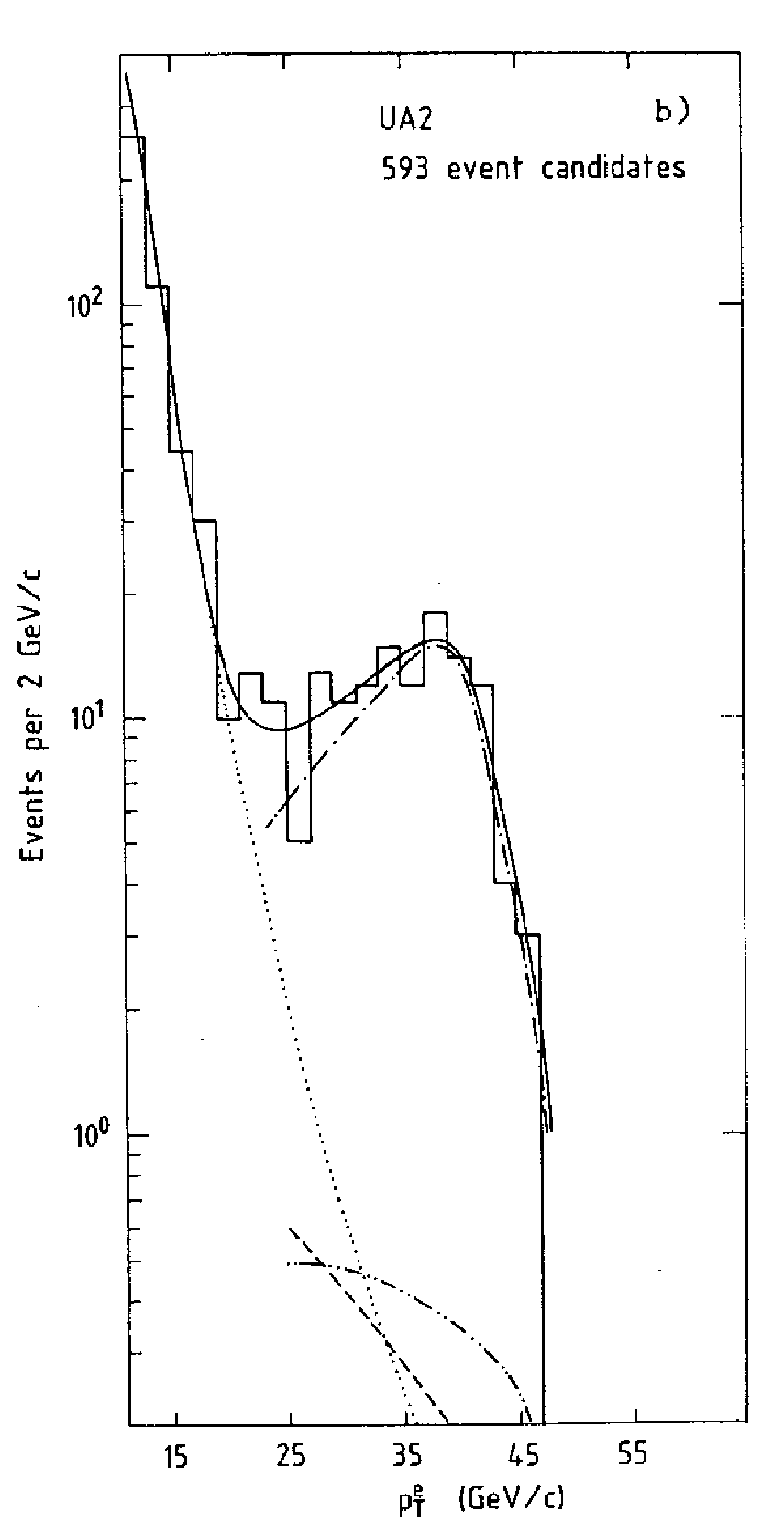,height=3.2in} 
\end{tabular}
\caption[]{ a) (Left) Predicted $p_T$ spectrum at $\sqrt{s}$=300 GeV 
from 
inclusive $\pi^0$, background $e^+$ from Dalitz decay of $\pi^0$, and $e^+$ 
from $W^+$ decay~\cite{FPMJT}. b) UA2 $e^{\pm}$ spectra with ``opposite azimuth'' cuts~\cite{UA2ZPC}.  
\label{fig:f1}}
\end{center}
\end{figure}

	During all the work in the early 1990's for the RHIC Spin Collaboration, with the prodding of Aki Yokosawa of Argonne National Laboratory and the help of Jacques Soffer~\cite{BS}, we figured out what we could learn about the proton spin structure by using the parity violating single spin asymmetry, $A_L$, in $W^+$ and $W^-$ production at RHIC. In fact, I made a presentation for the RSC at the PAC meeting of October 14, 1993 where I said that ``with PHENIX, I can find $W^{\pm}\rightarrow e^{\pm} +X$ in my sleep.'' I already knew this from the Snowmass '82 work, but in the intervening period, with Frank Paige~\cite{FPMJT} and others (Fig.~\ref{fig:f1}a), we had done some more work on the acceptance and beautiful  Jacobian peak in PHENIX because the $W$ at RHIC is essentially produced at rest. 
Also since UA2 had shown the method, practically in textbook form (Fig.~\ref{fig:f1}b)~\cite{UA2ZPC}, with the classic signature~\cite{Zichichi-Dubna} which had agreed with our Snowmass plots, I knew that we had in the PHENIX central arms more than the minimum needed to find a clean sample of $e^{\pm}$ from $W^{\pm}$ decays: a factor of 1000 charged hadron  rejection for $p_T>10$ GeV/c; precision EM Calorimetry out to 50 GeV; momentum resolution sufficient to resolve the charge of $e^{\pm}$ out to 50 GeV/c; a good trigger, as the $W$ is only $10^{-8}$ of the total cross section. 
 
\begin{figure}[!htb]
\begin{center}
\includegraphics[height=3.0in]{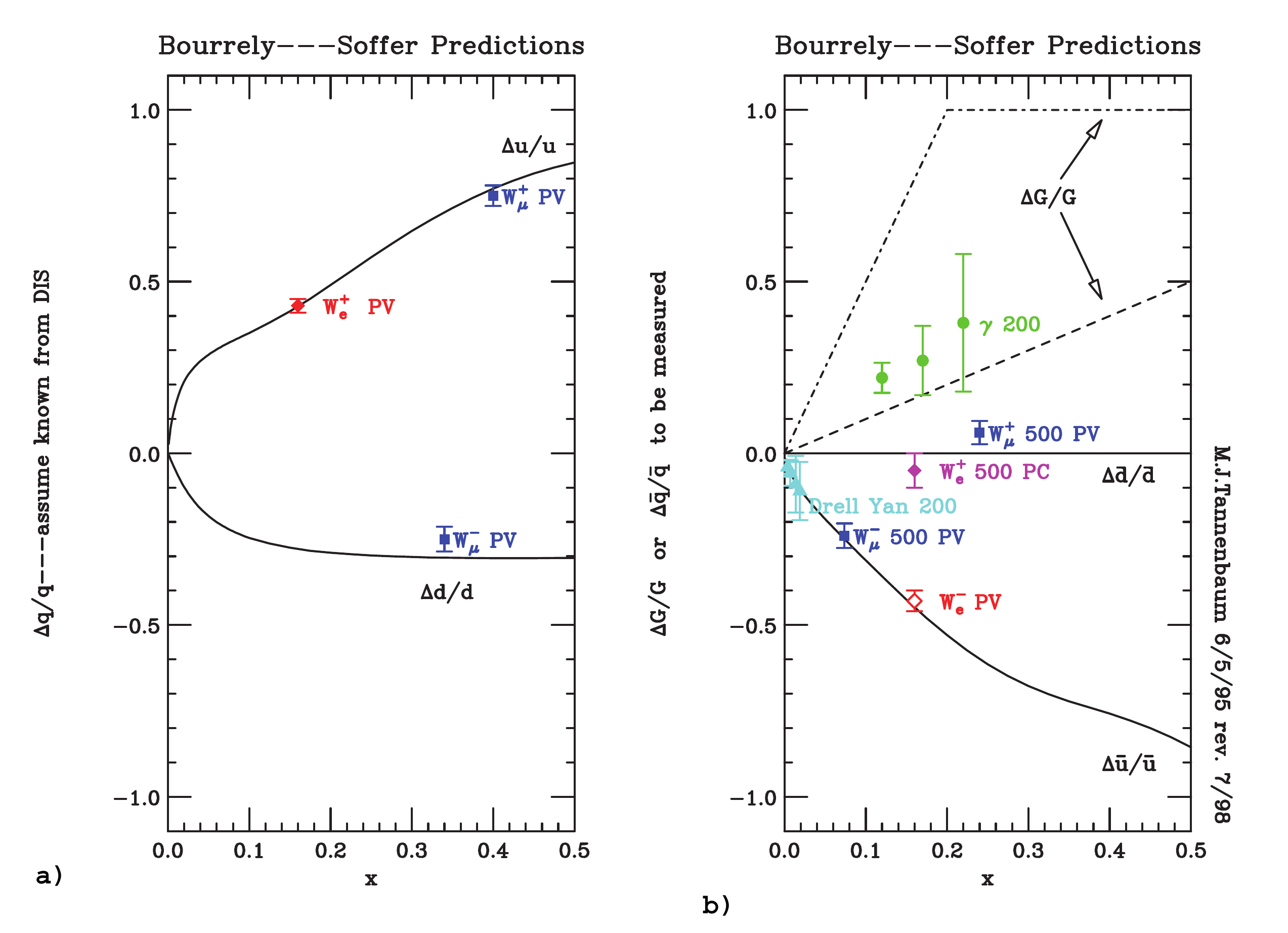}
\includegraphics[height=3.0in]{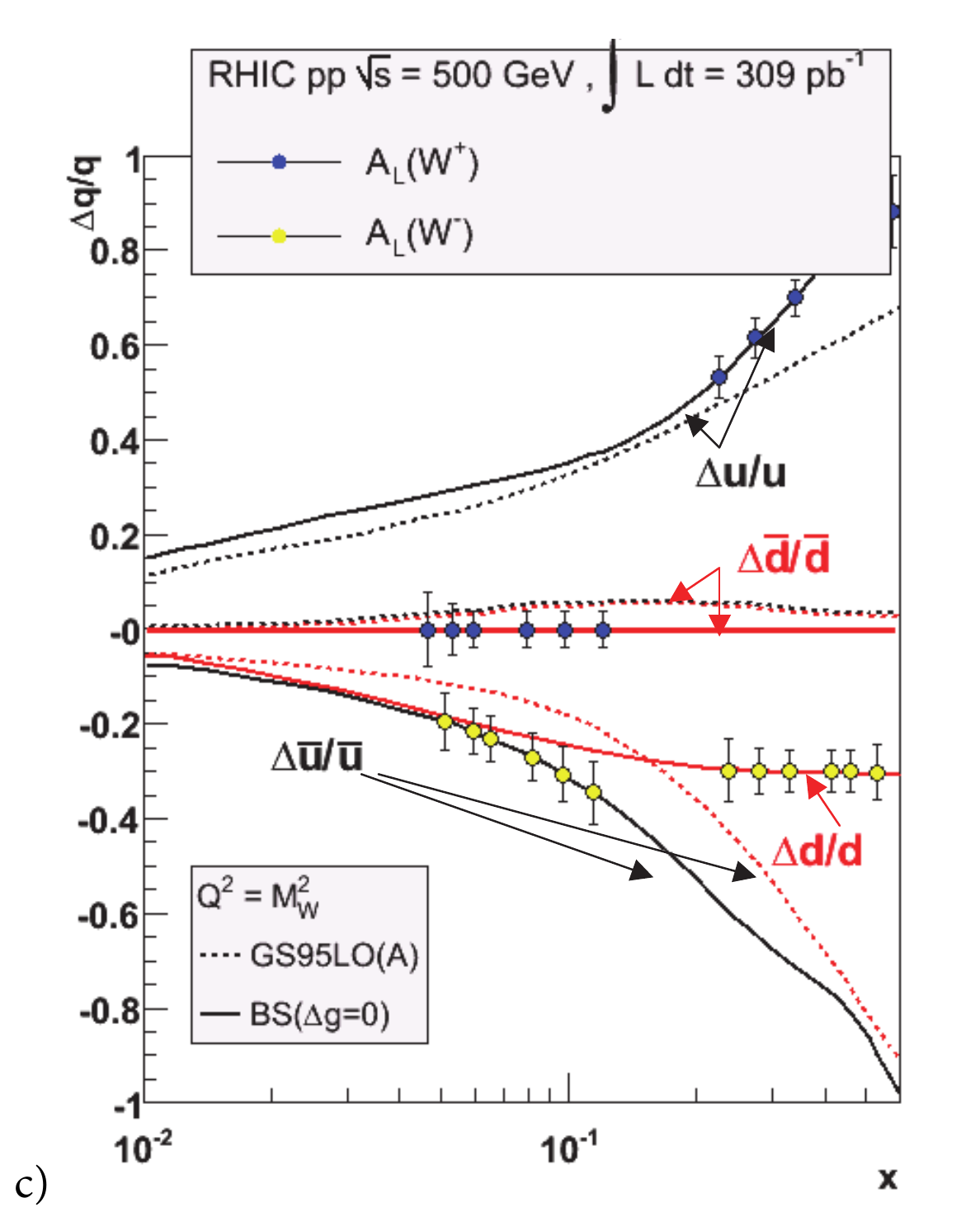}
\end{center}\vspace*{-0.12in}
\caption[]{a,b) Expected sensitivities for spin-structure function measurements in PHENIX at mid-rapidity $|y|<0.35$ shown with Bourrely-Soffer distributions~\cite{BS95} for 800 pb$^{-1}$ at $\sqrt{s}=500$ GeV and 320 pb$^{-1}$ at $\sqrt{s}=320$ GeV. c) Saito plot~\cite{Saito95}.}
\label{fig:Tannenbaumplot}\vspace*{-0.12in}
\end{figure}

For the final RHIC Spin Review in 1995, I made a plot of what precision to expect for the spin structure functions, $\Delta \bar{q}(x)/\bar{q}(x)$, of identified quarks---which are a unique feature of parity violating measurements because the $W$ is coupled to flavor not electric charge~\cite{BS95}---as well as the more standard gluon helicity density, $\Delta G(x)/G(x)$, with the proposed integrated luminosity and the PHENIX detector then under construction (Fig.~\ref{fig:Tannenbaumplot}a,b)~\cite{spin95}.
A similar plot was made by Naohito Saito for forward $W\rightarrow\mu(+\nu)$ detection $1.1<|y|<2.3$ at the same meeting (Fig.~\ref{fig:Tannenbaumplot}c). 

I also said a few other things at these meetings about why ``Parity Violation searches at RHIC satisfy all my criteria for the Maximum Discovery Potential'' that people still seem to remember:
\begin{itemize}
\item Look where most theorists predict that nothing will be found.
\item Look in a channel where the known rates from conventional processes are small, since low background implies high sensitivity for something new.
\item Be the first to explore a new domain---something that has never been measured by anybody else.
\end{itemize}

\begin{figure}[!hbt]
\begin{center}
\includegraphics[height=2.5in]{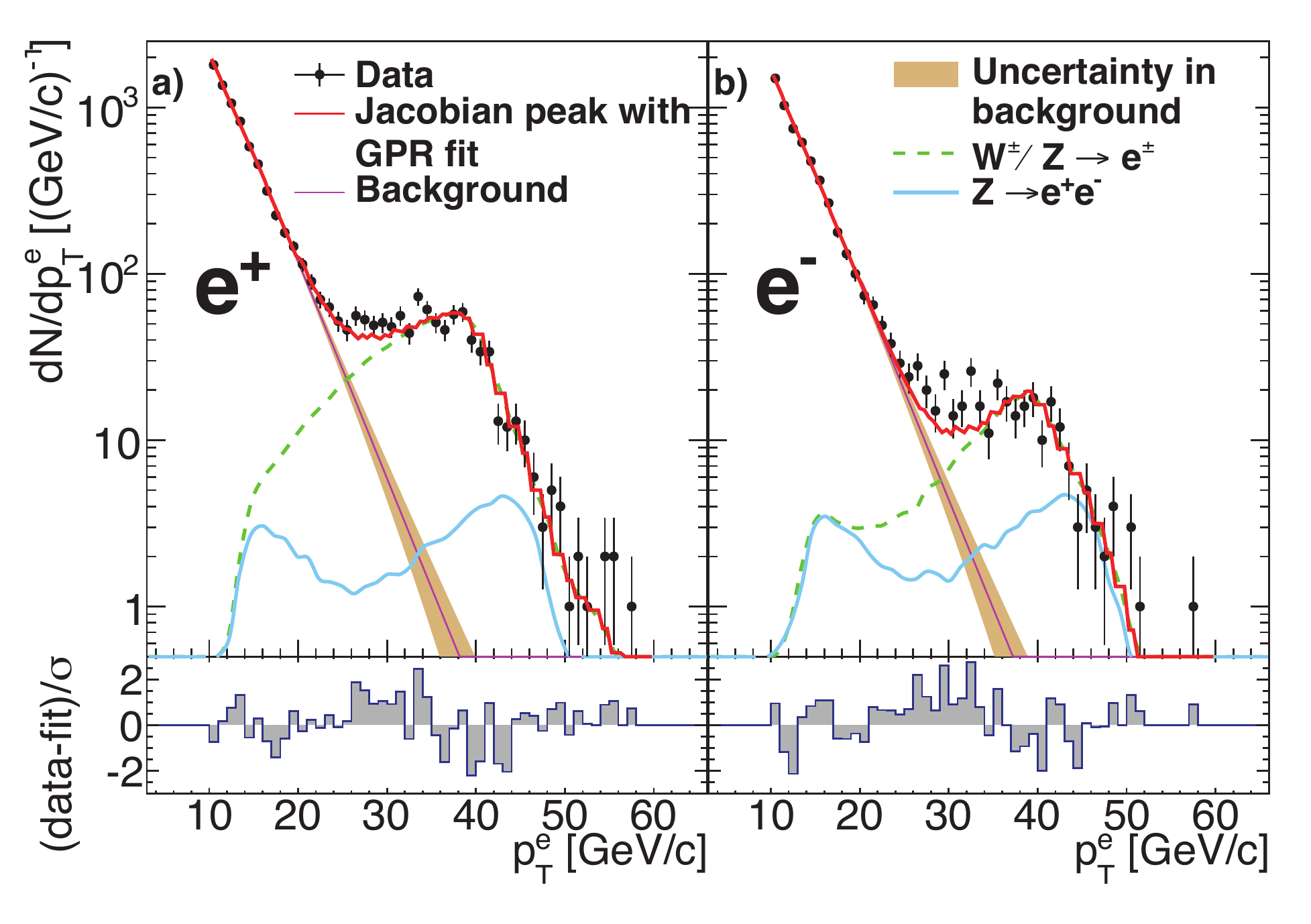}
\includegraphics[height=3.0in]{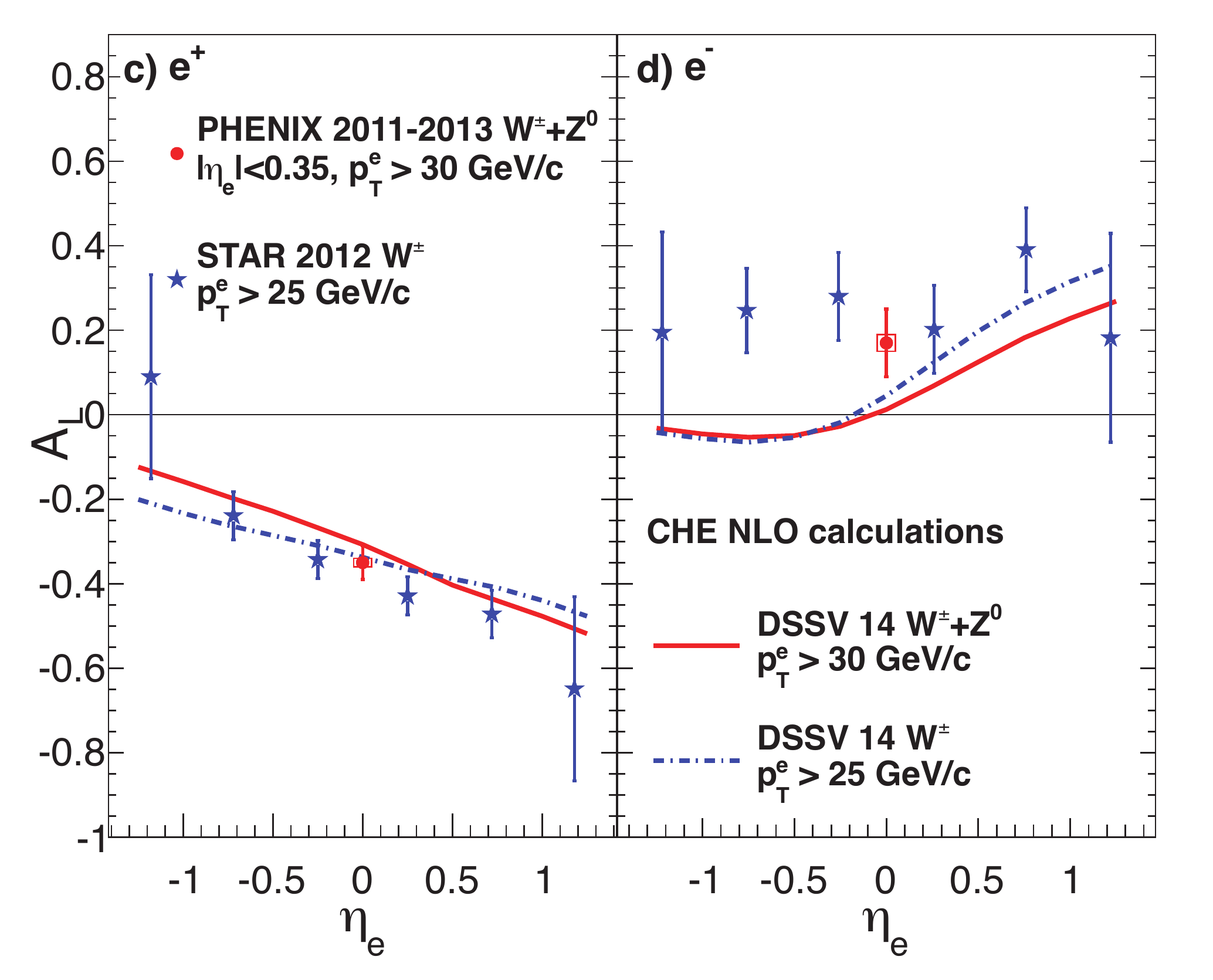}
\end{center}\vspace*{-0.12in}
\caption[]{a,b) PHENIX $W\rightarrow e^{\pm}+X$ measurements~\cite{PXW16}; c,d) PHENIX and STAR measurements of the parity violating longitudinal asymmetry $A_L$~\cite{PXW16} compared to modern predictions~\cite{DSSV14}. Note the different sign convention, $A_L$ is negative for the $W^+$, compared to Fig.~\ref{fig:Tannenbaumplot}. }
\label{fig:STARPXW}\vspace*{-0.12in}
\end{figure}

We actually had to wait a long time, more than 30 years since our original plots at Snowmass '82, to make beautiful $W$ measurements at RHIC, because it involved running the machine for polarized proton collisions at $\sqrt{s}=500$ GeV instead of the $\sqrt{s}=200$ GeV p$+$p data needed for comparison with the Au$+$Au collisions at $\sqrt{s_{NN}}=200$ GeV. The PHENIX~\cite{PXW16} and STAR~\cite{STARW14} measurements of the $W\rightarrow e^{\pm}+X$ spectrum in Fig.~\ref{fig:STARPXW}a,b nicely show the Jacobian peak as described by Zichichi~\cite{Zichichi-Dubna}, still valid after 50 years. The STAR and PHENIX measurements of the parity violating longitudinal single spin asymmetry $A_L$ for the detected $e^-$ and $e^+$ are shown in Fig.~\ref{fig:STARPXW}c,d as a function of the pseudorapidity $\eta$ of the $e^{\pm}$ and are in reasonable agreement with the present DSSV14~\cite{DSSV14} predictions for the parton spin asymmetries as well as the older predictions~\cite{BS95} in Fig.~\ref{fig:Tannenbaumplot}.
These are very nice long-awaited measurements but nothing that would be considered a discovery.

On the other hand we were actually the first to explore a new domain in Relativistic Heavy Ion Physics at RHIC, where we did in fact make a few discoveries, one of which was related to the Higgs. 
\subsection{The Higgs boson in RHI physics?}
\begin{figure}[!htb]
\begin{center}
\includegraphics[width=0.8\linewidth]{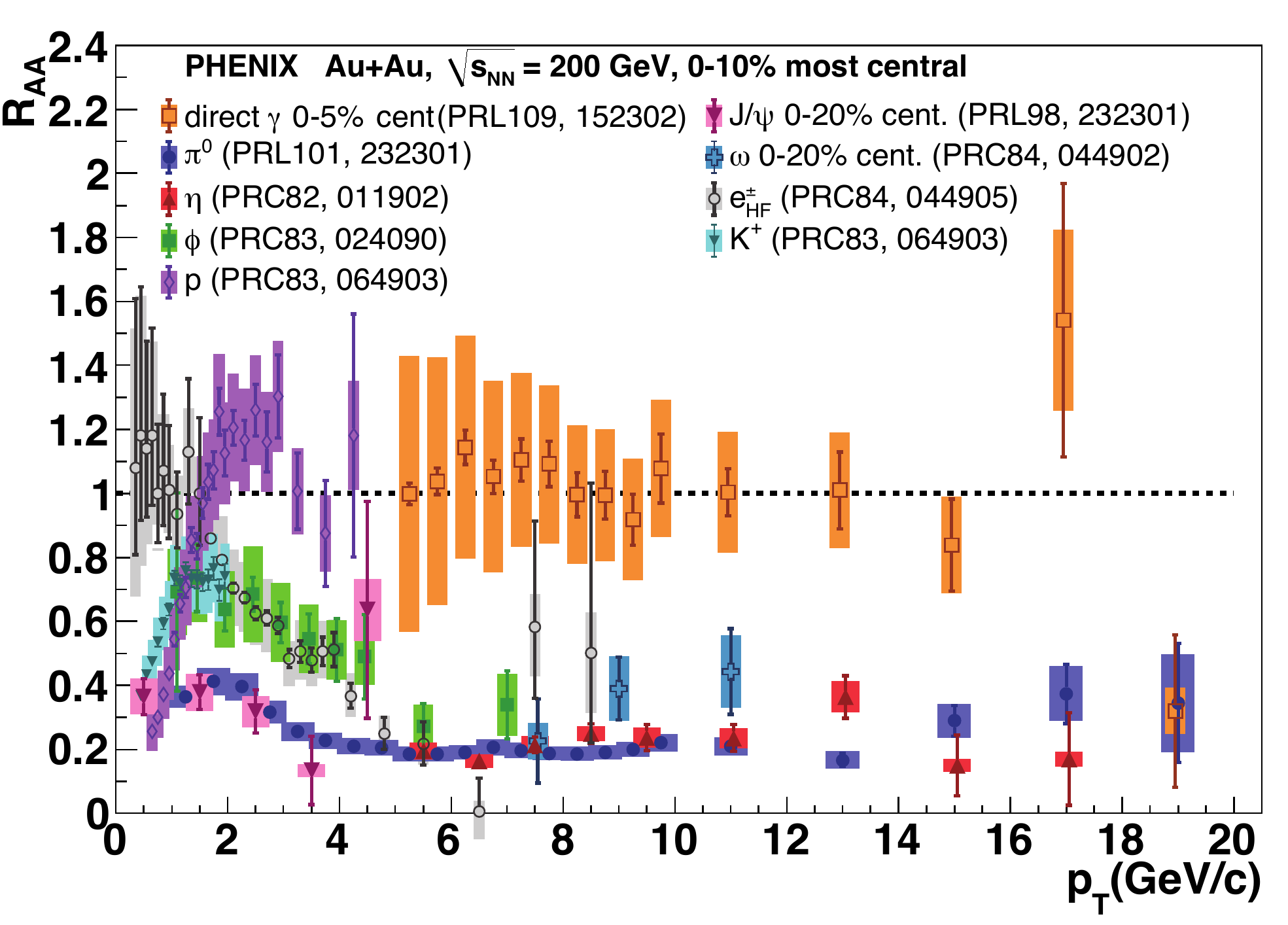}
\end{center}\vspace*{-0.12in}
\caption[]{PHENIX measurements of the suppression $R_{AA}$ of identified particles with references to publication as a function of transverse momentum $p_T$. }
\label{fig:Tshirt}\vspace*{-0.12in}
\end{figure}

The discovery at RHIC that particles produced with large transverse momenta $p_T$ from hard parton-parton scattering in the colliding Au$+$Au nuclei are suppressed in rate compared to the expected increase in cross section from p$+$p collisions for point-like scattering---a factor of $A$ larger in p$+$A and $A^2$ larger in A$+$A collisions---is arguably {\it the} major discovery in Relativistic Heavy Ion physics~\cite{ppg003}. In Figure~\ref{fig:Tshirt}, $R_{\rm AA}(p_T)$, the ratio of the measured to expected cross section in the most central (smallest impact parameter) Au$+$Au collisions as a function of $p_T$ is shown for the variety of identified particles measured by the PHENIX experiment.  

The striking differences of $R_{AA}(p_T)$ in central Au+Au collisions for the different particles in Fig.~\ref{fig:Tshirt} illustrates the importance of particle identification for understanding the physics of the medium produced at RHIC. Most notable are: the equal suppression of $\pi^0$ and $\eta$ mesons by a constant factor of 5 ($R_{AA}=0.2$) for $4\leq p_T \leq 15-20$ GeV/c; the equality of suppression of direct-single $e^{\pm}$ (from heavy quark $c$, $b$ decay) and $\pi^0$ at $p_T\gsim 5$ GeV/c; the non-suppression of direct-$\gamma$ for $p_T\geq 4$ GeV/c. For $p_T\gsim 4$ GeV/c, the hard-scattering region,  the fact that all hadrons are suppressed, but direct-$\gamma$ are not suppressed, indicates that suppression is a medium effect on outgoing color-charged partons likely due to energy loss by coherent Landau-Pomeranchuk-Migdal radiation of gluons, predicted in pQCD~\cite{BDMPS}, which is sensitive to properties of the medium.

The equality of suppression at $p_T\gsim 5$ GeV/c for $\pi^0$ from light quarks ($m\sim 4$ MeV) and direct-single $e^{\pm}$ from heavy quarks , where $b$ quarks ($m\gsim 4$ GeV) dominate, was a major discovery~\cite{ppg066}. It strongly disfavored the QCD energy-loss explanation of jet-quenching because, naively, the heavy $b$ quark should radiate much less than light quarks and gluons in the medium. On the other hand, it opened up a whole range of new possibilities including string theory (see Ref.~\cite{ppg066} for details).

\subsection{Zichichi to the rescue?}
  In September 2007,  I again happened to be reading an article by Antonino (Nino) Zichichi, ``Yukawa's gold mine'' in the CERN Courier taken from his talk at the 2007 International Nuclear Physics meeting in Tokyo, Japan, in which he proposed:``We know that confinement produces masses of the order of a giga-electron-volt. Therefore, according to our present understanding, the QCD colourless condition cannot explain the heavy quark mass. However, since the origin of the quark masses is still not known, it cannot be excluded that in a QCD coloured world, the six quarks are all nearly massless and that the colourless condition is `flavour' dependent.'' 
  
  Nino's idea really excited me even though, or perhaps because, it appeared to overturn two of the major tenets of the Standard Model since it seemed to imply that: QCD isn't flavor blind;  the masses of quarks aren't given by the Higgs mechanism.  Massless $b$ and $c$ quarks in a color-charged medium would be the simplest way to explain the apparent equality of gluon, light quark and heavy quark suppression indicated by the equality of $R_{AA}$ for $\pi^0$ and direct single-$e^{\pm}$ in regions where both $c$ and $b$ quarks dominate. Furthermore RHIC and LHC-Ions are the only place in the Universe to test this idea. 
  It may seem surprising that I would be so quick to take Nino's idea so seriously; but this confidence dates from my graduate student days as I discussed above when I had read Nino's method~\cite{Zichichi-Dubna} of detecting the $W$, which is exactly how it was discovered. I contacted him to discuss this new idea and he invited me to speak at his International School of Subnuclear Physics (ISSP) in Erice, Sicily, Italy.
   
  Nino's idea seemed much more reasonable to me than the string theory explanations of heavy-quark suppression (especially since they could't explain light-quark suppression). Nevertheless, just to be safe, I asked some distinguished theorists what they thought, with these results:
  \begin{itemize}
  \item Stan Brodsky:``Oh, you mean the Higgs field can't penetrate the QGP.''
 \item Rob Pisarski: `` You mean that the propagation of heavy and light quarks through the medium is the same.''
 \item Chris Quigg (Moriond 2008): ``The Higgs coupling to vector bosons $\gamma$, $W$, $Z$ is specified in the standard model and is a fundamental issue. One big question to be answered by the LHC is whether the Higgs gives mass to fermions or only to gauge bosons. The Yukawa couplings to fermions are put in by hand and are not required.'' ``What sets fermion masses, mixings?"
 \item Bill Marciano:``No change in the $t$-quark, $W$, Higgs mass relationship   if there is no Yukawa coupling: but there could be other changes.''
 \item Steve Weinberg: ``Lenny Susskind and I had a model, Technicolor (or Hypercolor),  that worked well in the vector boson sector but didn't give mass to the fermions.''
 \end{itemize}
	 
If Nino's proposed effect were true, that the masses of fermions are not given by the Higgs particle, and we could confirm the effect at RHIC or LHC-Ions, this would be a case where we Relativistic Heavy Ion Physicists may have something unique to contribute at the most fundamental level to the Standard Model, which would constitute a ``transformational discovery.'' Of course, I realized that the LHC could falsify this idea by finding the Higgs decay to $b-\bar{b}$ at the expected rate in p$+$p collisions. 

\begin{figure}[!hbt]
\begin{center}
\includegraphics[width=0.6\linewidth]{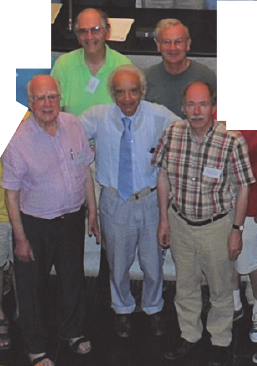}
\end{center}
\caption[]{Some attendees of ISSP50 on June 28, 2012 (Left to Right) Peter Higgs, MJT, Nino Zichichi, Frank Close, Gerard 't Hooft}
\label{fig:ISSP2012photo}\vspace*{-0.12in}
\end{figure}

At the 50th course of the ISSP from June 23--July 2,2012~\cite{ISSP50proc}, there was an unusually large number of invited distinguished guests, such as Peter Higgs (Fig.~\ref{fig:ISSP2012photo}) 	
and Murray Gell-Mann, because it was also the 40th Anniversary of QCD. I presented a talk about ``Highlights from BNL and RHIC'' on Wednesday June 27; then something very interesting happened.

In the question period after my talk, when I was asked about some theory calculation of the PHENIX data, I said that I didn't discuss theory, I leave that to the theorists. I added that ``I don't even talk about the Higgs boson until I see one ...'' (Strictly this was not true, as I did mention the Higgs boson because I doubted that it gives mass to fermions.) At dinner, I spoke to Peter Higgs and said that I hope he wasn't offended by my statement---he wasn't. Then he said,``You're from Brookhaven, right. Make sure to tell Sid Kahana that he was right about the top quark 175 GeV and the Higgs boson 125 GeV''~\cite{KahanaKahana}. I guess that it was no accident that the discovery of the Higgs boson at exactly that mass was announced on July 4, 2012 at CERN, just enough time for the distinguisted guests to make the trip from Erice.

In the past several years, the Higgs boson couplings have been measured with steadily increasing precision at the LHC~\cite{ATLAScoupling2016}. Nevertheless, even with all the excellent theoretical and experimental results, we still do not understand my PhD thesis problem: why the muon mass is 206.77 times larger than that of the electron. I wonder how much longer we will have to wait for that!

\section*{Acknowledgements}
 Research supported by U.S. Department of Energy, Contract No. DE-SC0012704.   

\end{document}